\documentclass[12pt,a4paper,fleqn]{article}

\usepackage[DIV12]{typearea}
\usepackage{amsmath}
\usepackage{amssymb}
\usepackage{amsfonts}
\usepackage{amscd}
\usepackage{graphicx}
\usepackage[footnotesize]{caption2}
\usepackage{cite}
\usepackage[x11names]{xcolor}
\usepackage{mathrsfs}
\usepackage{bbm}
\usepackage{psfrag}

\DeclareMathOperator{\tr}{Tr}

\newcommand{\eVdist}{\kern-0.06667em}

\newcommand{\ps}{\text{\sc ps}}
\newcommand{\GG}{\text{\sc gg}}

\newcommand{\sm}{\text{\sc sm}}

\def\beqra{\begin{eqnarray}}
\def\eeqra{\end{eqnarray}}

\def\beq{\begin{equation}}
\def\eeq{\end{equation}}

\def\beq{\begin{equation}}
\def\eeq{\end{equation}}
\def\bea{\begin{eqnarray}}
\def\eea{\end{eqnarray}}

\renewcommand{\[}{\left[}

\allowdisplaybreaks[1]

\begin{document}

\begin{titlepage}
\renewcommand{\thefootnote}{\alph{footnote}}

\begin{flushright}
DESY 08-098\\
TUM-HEP 712/09\\
SISSA 07/2009/EP
\end{flushright}

\vspace*{1.0cm}

\renewcommand{\thefootnote}{\fnsymbol{footnote}}

\begin{center}
{\LARGE\bf Enhanced Symmetries of Orbifolds\\ from Moduli Stabilization
} 

\vspace*{1cm}
\renewcommand{\thefootnote}{\alph{footnote}}

\textbf{
Wilfried Buchm\"uller\footnote[1]
{Email: \texttt{wilfried.buchmueller@desy.de}},
Riccardo Catena\footnote[2]{Email: \texttt{catena@sissa.it}}
\\and
Kai Schmidt-Hoberg\footnote[3]{Email: \texttt{kschmidt@ph.tum.de}}
}
\\[5mm]
{\it
$^a$Deutsches Elektronen-Synchrotron DESY, 22603 Hamburg, Germany\\
$^b$ISAS-SISSA, 34013 Trieste, Italy\\
$^c$Physik Department T30, Technische Universit\"at M\"unchen,\\ 85748 Garching, Germany
}

\end{center}

\vspace*{1cm}              

\begin{abstract}
\noindent 
We study a supersymmetric field theory in six dimensions compactified on
the orbifold ${T^2/\mathbbm Z}_2$ with two Wilson lines. After supersymmetry
breaking, the Casimir energy fixes the shape moduli at fixed points in field 
space where the symmetry of the torus lattice is enhanced. Localized 
Fayet-Iliopoulos terms stabilize the volume modulus at a size much smaller
than the inverse supersymmetry breaking scale. All moduli masses are smaller
than the gravitino mass. 
\end{abstract}
\end{titlepage}
\newpage

\section{Introduction}

Higher-dimensional theories provide a natural framework for extensions of the 
supersymmetric standard model which unify gauge interactions with gravity 
\cite{wit85}. In recent years, phenomenologically attractive examples 
have been constructed in five and six dimensions  compactified on orbifolds, 
and it has become clear how to embed such orbifold GUTs into the heterotic 
string \cite{rat07}. 

An important problem in orbifold compactifications is the stabilization
of moduli. In the following we study this question for an $\mathrm{SO(10)}$
model in six dimensions (6D) \cite{abc03,bks05} which, compared to models 
derived 
from the heterotic string \cite{bls07,bs08}, has considerably simpler field content.
The paper extends previous work which demonstrated that the compact dimensions
can be stabilized at small radii, $R \sim 1/M_{\mathrm{GUT}}$, much smaller 
than the inverse supersymmetry breaking scale $1/\mu$ \cite{bcs08}.

A crucial ingredient for the stabilization of compact dimensions is the
Casimir energy of bulk fields \cite{ac83}. Various aspects of the Casimir
energy for 6D orbifolds have already been studied in \cite{pp01,pp03,ghx05}.
Stabilization of the volume modulus can be
achieved by means of massive bulk fields, brane localized kinetic terms or
bulk and brane cosmological terms \cite{pp01}. Alternatively, the interplay 
of one- and two-loop contributions to the Casimir energy can lead to a 
stabilization at the length scale of higher-dimensional couplings \cite{gh05}.
Furthermore, fluxes and gaugino condensates play an important role 
\cite{bht06,hml08}. The mechanism studied in this paper is based on
expectation values $\mathcal{O}(M_{\rm GUT})$ of bulk fields, 
induced by local Fayet-Iliopoulos (FI) terms, and localized supersymmetry 
breaking leading to gaugino mediation \cite{kks99,clx99}.
In the framework of moduli mediation this mechanism is discussed in \cite{bms09}.

For a rectangular torus it has been shown in \cite{bcs08} that the interplay 
of `classical' and one-loop contributions to the vacuum energy density can 
stabilize the compact dimensions at $R \sim 1/M_{\rm GUT}$. Here we study
the stabilization of all three shape and volume moduli of the torus.
Remarkably, it turns out that the minimum occurs at a point with  `enhanced
symmetry', where the torus lattice
corresponds to the root lattice of $\mathrm{SO(5)}$.
Tori defined by Lie lattices are the starting point for orbifold 
compactifications in string theory, which lead to large discrete
symmetries \cite{knx06}. These restrict Yukawa couplings and can forbid or 
strongly suppress the $\mu$-term of the supersymmetric standard model 
\cite{bs08,knx08}. Enhanced discrete 
symmetries have previously been discussed in connection with string vacua \cite{din99}.

The paper is organized as follows. In Section~2 we discuss symmetries of the 
compact space and the associated moduli fields, whereas the relevant features 
of the
considered 6D orbifold GUT model are briefly described in Section~3. The 
Casimir energies of scalar fields with different boundary conditions are 
analyzed in Section~4. These results are the basis for the moduli stabilization
discussed in Section~5.
The Appendix deals with the evaluation of Casimir sums.

\section{Modular Symmetries of Orbifolds}

In this section we briefly discuss the geometry of the compact space
and the associated three moduli fields. The torus $T^2$, 
and also the $T^2/\mathbbm{Z}_2$ orbifold,
can be parameterized by the volume parameter $\mathcal{A}$ and 
the complex shape parameter $\tau=\tau_1+i \tau_2$.
Following \cite{pp01}, we choose the following metric for $M^4\times T^2$,
\beq
\label{background6D}
ds^2 = \mathcal{A}^{-1}~g_{\mu \nu} dx^\mu dx^\nu +
\mathcal{A}~\gamma_{ij} dy^i dy^j \ ,
\eeq
where $y^i \in [0,L]$, and the metric $\gamma_{ij}$ on the torus is 
given by
\beq
\gamma_{ij} = \frac{1}{\tau_2}
\left( \begin{array}{cc} 1 & \tau_1 \\\tau_1 & |\tau|^{2} \end{array}\right)\ .
\eeq
4D Minkowski space corresponds to $g_{\mu \nu} = \eta_{\mu \nu}$, and the
induced metric at the orbifold fixed points is 
$\tilde{g}_{\mu \nu} = \mathcal{A}^{-1}\eta_{\mu \nu}$.
The kinetic terms of the moduli fields are obtained by dimensional reduction
from the 6D Einstein-Hilbert action, 
\begin{align}
S &=  \frac{M_6^4}{2} \int d^6 x \sqrt{G}R(G)  \nonumber\\
&=   \frac{M^{4}_6 L^2}{2}\int  d^4 x \sqrt{g}  \left( R(g) 
+ \frac{g^{\mu\nu} \partial_{\mu} \mathcal{A}  
\partial_{\nu} \mathcal{A}}{\mathcal{A}^2} +
\frac{g^{\mu\nu} \partial_{\mu} \tau  
\partial_{\nu} \tau^*}{2\mathcal{\tau}_2^2} \right)\ .  
\label{action}
\end{align}
Here $\tau^* = \tau_1-i \tau_2$, and the Ricci scalar $R$ is 
evaluated with the metric indicated in parenthesis. Note, that 
$M_6^2 L$ is not the 4D Planck mass, since $g_{\mu\nu}$ does not determine the physical
distance in the non-compact dimensions, cf.~(\ref{background6D}). Once the area modulus 
$\mathcal{A}$ is stabilized at $\mathcal{A}_0$, a constant Weyl rescaling  
$g_{\mu\nu} = \mathcal{A}_0 \bar{g}_{\mu\nu}$
yields the physical Planck mass
$M_4=\sqrt{\mathcal{A}_0L^2} M_6^2$, with $\mathcal{A}_0 L^2$ being the 
area of the torus. 

In this paper we extend previous work \cite{bcs08}, where only a rectangular 
torus lattice was considered. 
Note, that the torus can alternatively be described
by the two radii $R_{1,2}$ of the torus lattice and the 
angle $\theta$ between them.
The relation between the two sets of parameters is given by
\begin{align}
2\pi R_1&= L \sqrt{\frac{\mathcal{A}}{\tau_2}} \ , \quad
2\pi R_2= |\tau|L \sqrt{\frac{\mathcal{A}}{\tau_2}} \ , \quad
\theta=\text{arccos}\frac{\tau_1}{|\tau|}\ .
\end{align}
The rectangular torus in \cite{bcs08} has been parameterized in terms of the 
two radii $R_{1,2}$, corresponding to $\tau_1=0$ and $\tau_2=R_2/R_1$.

\begin{figure}
  \centering
    \includegraphics[width=7.8cm,height=6cm]{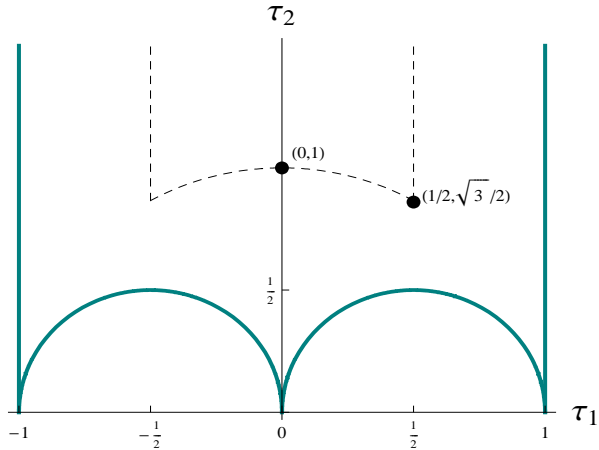}  
  \caption{Fundamental domain for the modular groups $SL(2,\mathbbm{Z})$ and $\Gamma(2)$.}
  \label{fig:fund}
\end{figure}

The group $\mathrm{SL}(2,\mathbbm{Z})$ of modular transformations 
\begin{align}
\label{modsym}
\tau\rightarrow \frac{a\tau+b}{c\tau+d} \ , \quad
a,b,c,d\in\mathbbm{Z}\ , \quad  ad-bc=1\ ,
\end{align}
relates modular parameters of
diffeomorphic tori. Distinct tori have modular parameters $\tau$ taking 
values in the fundamental region
$|\tau|\ge 1$, $-1/2\le \tau_1 \le 1/2$ and $\tau_2 > 0$ 
(cf.~Figure~\ref{fig:fund}).

The Kaluza-Klein mode expansion of bulk fields on the torus can be 
written as 
\begin{align}
\Phi(x,y)
&=\frac{1}{\sqrt{\mathcal{A}}L}\sum_{m,n=-\infty}^{\infty}\phi_{m,n}(x)
\exp\left\{\frac{2\pi i}{L \sqrt{\mathcal{A}\tau_2}} \left[m\left(\tau_2y_1-\tau_1 y_2\right) 
+n y_2\right] \right\} \ ,
\end{align}
with the corresponding Kaluza-Klein (KK) masses 
\begin{align}
\mathcal{M}^2_{m,n} 
&= \frac{(2\pi)^2}{\mathcal{A}L^2 \tau_2}|m\tau-n|^2 \ .
\end{align}
Note that the sum over all KK modes is modular invariant: the transformations 
associated with the two $\mathrm{SL}(2,\mathbbm{Z})$ generators,
$\tau \rightarrow \tau+1$ and $\tau\rightarrow-1/\tau$, correspond to
the relabeling of terms $(m,n)\rightarrow (m,n-m)$ and
$(m,n)\rightarrow (-n,m)$, respectively.

In the case of non-zero Wilson lines the KK masses take the
values \cite{ghx05}
\begin{align}
\label{eq:kkmasses}
\mathcal{M}^2_{m,n} &= \frac{(2\pi)^2}{\mathcal{A}L^2 \tau_2}|n+\beta-\tau(m+\alpha)|^2 \ , 
\end{align}
where $(\alpha,\beta)$ are real numbers.
For a $T^2/\mathbbm{Z}_2$ orbifold, $\alpha$ and $\beta$ are 
restricted, $\alpha,\beta \in\{0,1/2\}$.
The modular transformation (\ref{modsym}),
with $\tau_2 \rightarrow \tau_2/(|c\tau+d|^2)$, now corresponds to the
relabeling of KK modes
\begin{align}
\label{eq:kktrafo}
m+\alpha &\rightarrow a(m+\alpha)-c(n+\beta)  \ , \quad
n+\beta \rightarrow d(n+\beta)-b(m+\alpha) \ .
\end{align} 
Depending on the values of the discrete Wilson lines, the sum over KK modes is 
invariant under the full modular group $\mathrm{SL}(2,\mathbbm{Z})$ or some 
subgroup \cite{spa91}.
For $\alpha=\beta=0$, the Wilson lines are zero and 
$\mathrm{SL}(2,\mathbbm{Z})$ remains unbroken. In the case
$\alpha=0$ and $\beta=1/2$, modular invariance yields the additional 
restriction $c=0 \mod 2$ and $d=1 \mod 2$. Correspondingly,
for $\alpha=1/2$ and $\beta=0$ one finds the  restriction 
$a=1 \mod 2$ and $b=0 \mod 2$, while for $\alpha=\beta=1/2$ one has
$a,d=1 \mod 2$, $b,c=0 \mod 2$ or $a,d=0 \mod 2$ and $b,c=1 \mod 2$.
The largest common subgroup corresponds to $a,d=1 \mod 2$ and $b,c=0 \mod 2$,
which corresponds to $\Gamma(2)$ \cite{ran77}. 
The fundamental domain of the groups $\Gamma(2)$ and 
$\mathrm{SL}(2,\mathbbm{Z})$ are compared in Figure~\ref{fig:fund}.

We are interested in fixed points of the modular group in the upper half plane,
because the effective potential $V(\tau_1,\tau_2)$ has extrema at these
fixed points. To this end notice that a matrix 
$M \in \text{SL}(2,\mathbbm{Z}), M\neq \pm \mathbbm{1}$, has
a fixed point in the upper half plane if and only if $\tr M <2$.
This can be seen from the fixed point equation $Mz=z$ which implies
\begin{align}
& cz^2+(d-a)z-b=0 \ .
\end{align} 
Using the property $ad-bc=1$, one obtains for the solutions of this equation 
\begin{align}
\label{eq:fp}
z&=\frac{a-d\pm\sqrt{(a+d)^2-4}}{2c} \ .
\end{align}
We see that only for $(a+d)^2<4$ we have complex solutions in the upper half 
plane, whereas for $(a+d)^2 \ge 4$ there are only real solutions.
Clearly, only points on the edge of the fundamental domain can be fixed points,
because points within the fundamental domain are inequivalent and therefore 
cannot be mapped onto each other by a modular transformation.

It is well known that $\mathrm{SL}(2,\mathbbm{Z})$ has two fixed points at 
$(\tau_1,\tau_2)=(0,1)$ and $(\tau_1,\tau_2)=(1/2,\sqrt{3}/2)$, respectively.
For the case $c=0 \mod 2$ and $d=1 \mod 2$ there is a fixed point at 
$(\tau_1,\tau_2)=(1/2,1/2)$, while for $a=1 \mod 2$ and $b=0 \mod 2$ there is 
a fixed point at $(\tau_1,\tau_2)=(1,1)$.
Finally, in the case  $a,d=0 \mod 2$ and $b,c=1 \mod 2$ there is a fixed point
at $(0,1)$.
The subgroup $\Gamma(2)$ has no fixed points in the upper half plane.

\section{An Orbifold GUT Model}
\label{model}
As an example, we consider a 6D $\mathcal{N}=1$ $\mathrm{SO}(10)$ gauge theory 
compactified on an orbifold $T^2/{\mathbbm Z}_2^3$,
corresponding to $T^2/{\mathbbm Z}_2$ with two Wilson lines \cite{abc03}. 
The model has four inequivalent fixed points (`branes') with the unbroken 
gauge groups $\mathrm{SO}(10)$, the Pati-Salam group
$\mathrm{G}_{\ps}={\mathrm{SU}(4)}\times \mathrm{SU}(2) \times 
\mathrm{SU}(2)$, the extended Georgi-Glashow
group $\mathrm{G}_{\GG}= \mathrm{SU}(5)\times \mathrm{U}(1)_X$ and flipped 
$\mathrm{SU}(5)$, $\mathrm{G}_{\text{fl}}= \mathrm{SU}(5)'\times 
\mathrm{U}(1)'$, respectively. The intersection of 
these GUT groups yields the standard model group with an additional 
$\mathrm{U}(1)$ factor, 
$\mathrm{G}_{\sm}'= \mathrm{SU}(3)_C\times \mathrm{SU}(2)_L \times 
\mathrm{U}(1)_Y\times \mathrm{U}(1)_{X}$, 
as unbroken gauge symmetry below the compactification scale. 

The model has three {\bf 16}-plets of matter fields, localized at the 
Pati-Salam, the Georgi-Glashow, and the flipped $\mathrm{SU}(5)$ branes. 
Further,
there are two {\bf 16}-plets, $\phi$ and $\phi^c$, and two {\bf 10}-plets,
$H_5$ and $H_6$, of bulk matter fields. Their mixing with the brane fields 
yields the characteristic flavor structure of the model \cite{abc03}.

The Higgs sector consists of two {\bf 16}-plets, $\Phi$ and $\Phi^c$, 
and four {\bf 10}-plets, $H_1,\ldots, H_4$, of bulk hypermultiplets.
The hypermultiplets $H_1$ and $H_2$ contain
the two Higgs doublets of the supersymmetric standard model as zero modes,
whereas the zero modes of $H_3$ and $H_4$ are color triplets.
The zero modes of the {\bf 16}-plets are singlets and color triplets,
\begin{align}
&\Phi:\;\; N^c,\ D^c\;;\qquad \Phi^c:\;\; N,\ D\;.
\end{align}
The color triplets $D^c$ and $D$, together with the zero modes of $H_3$ and
$H_4$, acquire masses through brane couplings.

Equal vacuum expectation values of $\Phi$ and $\Phi^c$ form a flat direction 
of the classical potential,
\begin{align}\label{vev}
&\langle \Phi \rangle = \langle N^c \rangle 
= \langle N \rangle = \langle \Phi^c \rangle \;.
\end{align} 
Non-zero expectation values can be enforced by a brane superpotential term or 
by an FI-term localized at the GG-brane where the $\mathrm{U}(1)$ factor 
commutes with the standard model gauge group.

The expectation values (\ref{vev})
break $\mathrm{SO}(10)$ to $\mathrm{SU}(5)$, and therefore also
the additional $\mathrm{U}(1)_X$ symmetry, leading to bulk 
masses\footnote{For more 
details concerning the parity assignments and gauge 
symmetry breaking, see \cite{bcs08}.}
\begin{equation}\label{higgsmass}
\mathcal{M}^2 \simeq  g_6^2 \langle \Phi^c \rangle^2\;,
\end{equation}
where $g_6$ is the 6D gauge coupling, which is related to the 4D gauge
coupling by a volume factor, $g_4=g_6/\sqrt{\mathcal{A}L^2}$.

Supersymmetry breaking is naturally incorporated via gaugino mediation 
\cite{bks05}. The non-vanishing $F$-term of a brane field $S$ generates
mass terms for vector- and hypermultiplets. In the considered model, $S$
is localized at the $\mathrm{SO}(10)$ preserving brane, which yields the 
same mass for all members of an $\mathrm{SO}(10)$ multiplet. For the 
${\bf 45}$ vector multiplet
and the ${\bf 10}$ and {\bf 16} hypermultiplets of the Higgs sector one has
\begin{eqnarray}\label{mass}
\Delta S &=& \int \text{d}^4x \text{d}^2y\ \sqrt{\widetilde{g}}\ \delta^2(y) \left\{\int \text{d}^2 \theta  
 \frac{h}{2\Lambda^3} S {\rm Tr}[W^\alpha W_\alpha] + {\rm h.c.} 
 \right. \nonumber\\ 
&& \left. + \int \text{d}^4\theta \left(
\frac{\lambda}{\Lambda^4} S^\dagger S 
  \left(H_1^\dagger H_1 + H_2^\dagger H_2\right)  
 + \frac{\lambda'}{\Lambda^4} S^\dagger S 
  \left(H_3^\dagger H_3 + H_4^\dagger H_4\right)\right.\right.\nonumber\\ 
&& \hspace{16mm} \left.\left. + \frac{\lambda''}{\Lambda^4} S^\dagger S 
  \left(\Phi^\dagger \Phi + \Phi^{c\dagger} \Phi^c\right) \right)\right\}\;.
\end{eqnarray}
Here $\widetilde{g}_{\mu\nu}$ is the metric induced at the fixed point, and 
$W^\alpha(V)$, $H_1,\ldots, H_4$, $\Phi, \Phi^c$ are the 4D 
$\mathcal{N}=1$ multiplets
contained in the 6D $\mathcal{N}=1$ multiplets, which have positive parity
at $y=0$; $\Lambda$ is the UV cutoff of the model, which is much larger 
than the inverse size of the compact dimensions.
For the zero modes, the corresponding gaugino and scalar masses are given by
\begin{equation}\label{masses}
m_g = \frac{h\mu}{\mathcal{A} L^2\Lambda^2}\;, \quad 
m^2_{H_{1,2}} = - \frac{\lambda \mu^2}{\mathcal{A} L^2\Lambda^2}\;, \quad
m^2_{H_{3,4}} = - \frac{\lambda' \mu^2}{\mathcal{A}L^2\Lambda^2}\;, \quad
m^2_{\Phi} = - \frac{\lambda'' \mu^2}{\mathcal{A}L^2\Lambda^2}\;,  
\end{equation} 
where $\mathcal{A} L^2$ is the volume of the compact dimensions and
$\mu=F_S/\Lambda$. Note that the gaugino mass is stronger volume 
suppressed than the scalar masses.
This implies that the contribution of the vector multiplet to the Casimir energy relative to the
one of the hypermultiplets is also suppressed, as shown in the appendix.

\section{Casimir Energy on $T^2/\mathbbm{Z}_2^3$}

The Casimir energy of a real scalar field on the given orbifold background 
can be written as
\begin{align}\label{effpot}
V_M
&= \frac{1}{2} \left[
\sum \right]_{m,n}
\int \frac{\text{d}^4k_E}{(2\pi)^4} 
\log\left(k_E^2 +\frac{\mathcal{M}^2_{m,n}}{\mathcal{A}}+\frac{M^2}{\mathcal{A}}\right) 
\;,
\end{align}
with $\left[\sum \right]_{m,n}$ shorthand for the double sum
and $\mathcal{M}^2_{m,n}$ denoting the Kaluza-Klein masses, which are given by
(\ref{eq:kkmasses}) except for a factor of four due to the two additional 
$\mathbbm{Z}_2$ symmetries, which have been modded out.
The mass $M$ stands for bulk and brane mass terms.

The expression (\ref{effpot}) for the Casimir energy is divergent. Following
\cite{pp01}, we extract a finite piece using zeta function 
regularization, 
\begin{align}
V
&= - \frac{\text{d}\zeta(s)}{\text{d}s}\bigg|_{s=0} \;,
\end{align}
where 
\begin{align}
\hspace{-4mm}
\zeta(s) &= 
\frac{1}{2} \left[
\sum \right]_{m,n} \mu_r^{2s}
\int \frac{\text{d}^4k_E}{(2\pi)^4} 
\left(k_E^2 +  \frac{4(2\pi)^2}{\mathcal{A}^2 L^2\tau_2}|n+\beta-\tau(m+\alpha)|^2   +   \frac{M^2}{\mathcal{A}}\right)^{-s}\;.
\end{align}
Note that, as in dimensional regularization, a mass scale $\mu_r$ is 
introduced. The momentum integration can be performed, which yields 
\begin{align}\label{zeta}
\zeta(s)
&= \frac{1}{2} \frac{1}{(2\pi)^4} 
\pi^2 \frac{\Gamma(s-2)}{\Gamma(s)} \left[\sum \right]_{m,n}\mu_r^{2s}
\left( \frac{4(2\pi)^2}{\mathcal{A}^2L^2\tau_2}|n+\beta-\tau(m+\alpha)|^2    
+   \frac{M^2}{\mathcal{A}}\right)^{2-s} \nonumber \\
&= \frac{\mu_r^{2s}4^{2-s}(2\pi)^{-2s}\pi^2}
{2 \mathcal{A}^{4-2s}L^{4-2s}\tau_2^{2-s}(s-2)(s-1)} 
\nonumber \\ 
& \quad \left[\sum \right]_{m,n}
\left[(n+\beta-(m+\alpha)\tau_1)^2 + (m+\alpha)^2 \tau_2^2
+  \frac{\mathcal{A}L^2\tau_2}{4(2\pi)^2}  M^2\right]^{2-s}  \;.
\end{align}
\begin{figure}
  \centering
  \begin{minipage}[b]{7.8 cm}
    \includegraphics[width=7.8cm,height=6cm]{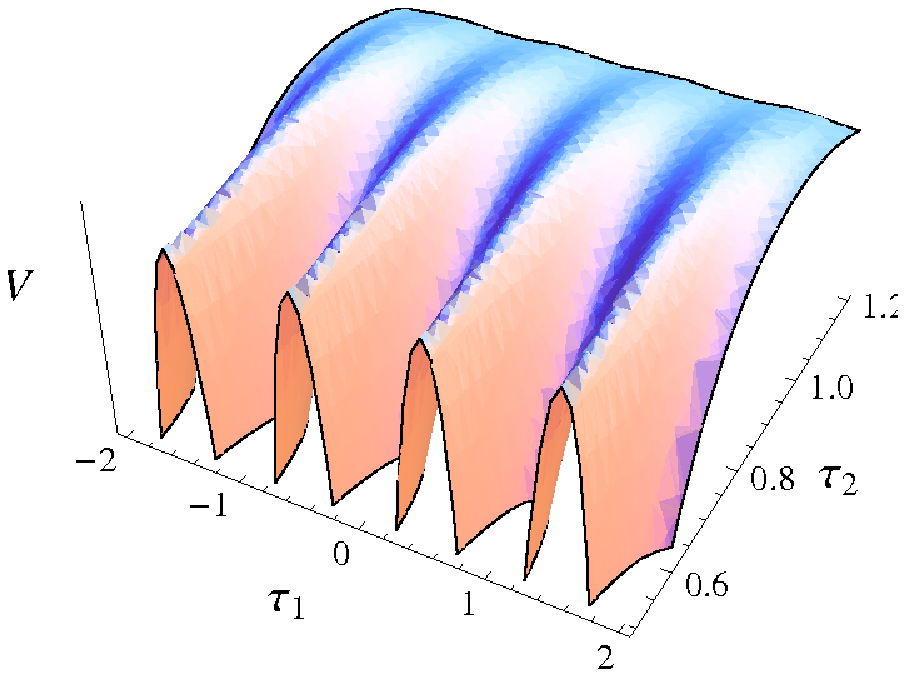}  
  \end{minipage}\hspace*{1cm}
  \begin{minipage}[b]{7.8 cm}
    \includegraphics[width=7.8cm,height=6cm]{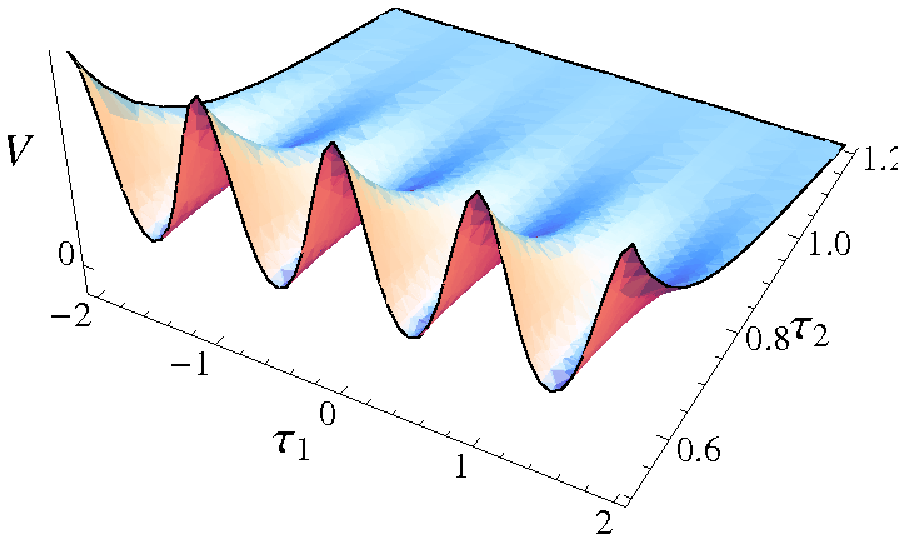}  
  \end{minipage}\vspace*{1cm}
  \begin{minipage}[b]{7.8 cm}
    \includegraphics[width=7.8cm,height=6cm]{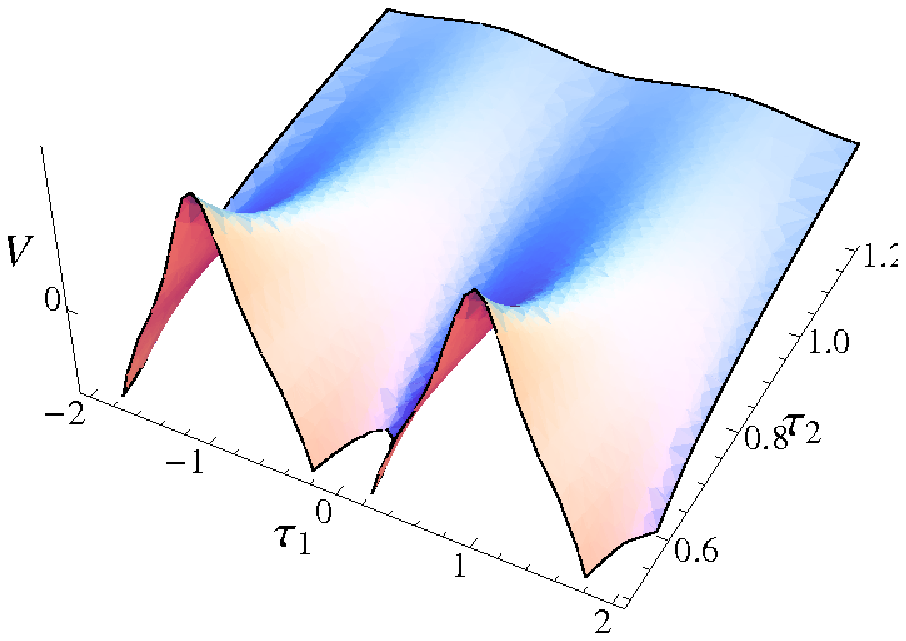}  
  \end{minipage}\hspace*{1cm}
  \begin{minipage}[b]{7.8 cm}
    \includegraphics[width=7.8cm,height=6cm]{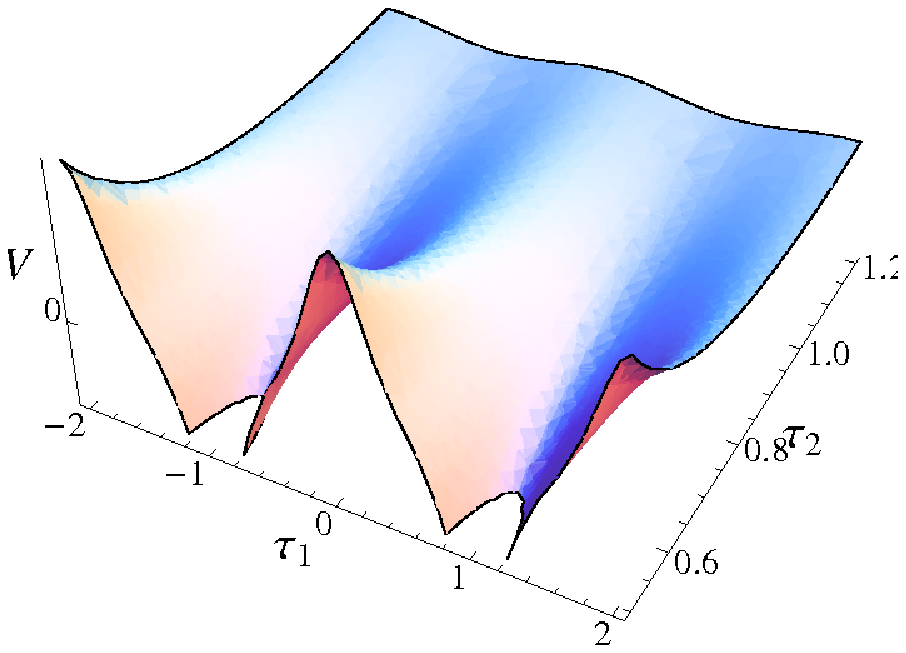}  
  \end{minipage}
  \caption{The different contributions to the Casimir energy for the different
           boundary conditions. Note that the potential is periodic in
           $\tau_1$ with period 1 for $\alpha=0$ and period 2 for $\alpha=1/2$.}
  \label{fig:shape}
\end{figure}
Carrying out the summations (cf.~Appendix) we find for the Casimir energy
\begin{align} 
\hspace{-6mm}V^{\alpha,\beta}_M 
=&\frac{M^6 L^2}{3072\pi^3 \mathcal{A}} \left[\frac{11}{12} 
-\log\left(\frac{M}{\sqrt{\mathcal{A}}\mu_r}\right)\right]
\nonumber \\
&- \frac{M^4}{64\pi^2\mathcal{A}^2 }
\left[\frac{3}{4}-\log\left(\frac{M}{\sqrt{\mathcal{A}}\mu_r}\right)\right]
\delta_{\alpha 0}\delta_{\beta 0}
\nonumber \\
&- \frac{M^3 \tau_2^{3/2}}{4 \pi^3\mathcal{A}^{5/2}L} 
\sum_{p=1}^{\infty}\frac{\cos(2\pi p \alpha)}{p^3}
K_{3}\left(p \tfrac{\sqrt{\mathcal{A}}L M}{2\sqrt{\tau_2}}\right) 
 \nonumber \\
&-\frac{32}{\mathcal{A}^4 L^4\tau_2^2} \sum_{p=1}^{\infty} 
 \sum_{m=0}^{\infty}\frac{1}{2^{\delta_{\alpha 0}\delta_{m 0}}}
\frac{\cos(2 \pi p (\beta-(m+\alpha)\tau_1))}{p^{5/2}} 
\nonumber \\
& \times\left(\tau_2^2(m+\alpha)^2 + \frac{\mathcal{A}L^2 \tau_2 M^2}{(4\pi)^2}  \right)^{\frac{5}{4}} 
 K_{5/2} \left(2\pi \,p \,\sqrt{\tau_2^2(m+\alpha)^2 +  \frac{\mathcal{A}L^2\tau_2 M^2}{(4\pi)^2}   }\right) . 
\end{align}
The different contributions to the Casimir energy are displayed in 
Figure~\ref{fig:shape} as function of the shape moduli $\tau_1$ and $\tau_2$
for fixed volume modulus $\mathcal{A}$.  

In supersymmetric theories there is a cancellation between bosonic and 
fermionic contributions, and the expression for the
Casimir energy is given by
\begin{eqnarray}\label{susy}
V&=&
  A \left( V_{M'}^{0,0}- V_M^{0,0}\right)
+ B \left( V_{M'}^{0,1/2}- V_M^{0,1/2}\right) 
\nonumber \\
&&+C \left( V_{M'}^{1/2,0} -V_M^{1/2,0}\right)
+ D \left( V_{M'}^{1/2,1/2}-V_M^{1/2,1/2}\right)\;,
\end{eqnarray}
where $M'=\sqrt{M^2+m^2}$, with supersymmetric mass $M$ and  
supersymmetry breaking mass $m$; the coefficients $A$,$B$,$C$,$D$ depend 
on the field content of the model. 
Note that even in the supersymmetric framework there are divergent bulk and
brane terms, which are proportional to the supersymmetry breaking  mass $m^2$,
unlike the case in Scherk-Schwarz breaking.
These divergencies have to be subtracted from the unrenormalized Casimir
energy to obtain a finite result,
and to tune the four-dimensional cosmological constant to zero.

\section{Stabilization}
 
\subsection{Shape Moduli}

Before discussing moduli stabilization for our particular orbifold GUT model, 
it is instructive to consider the shape moduli potential for varying field 
content, i.e., for different coefficients $A$,$B$,$C$,$D$. The modular 
symmetries (\ref{modsym}) of the four different contributions are given in 
Table~\ref{tab:modsym}.
\begin{figure}
  \centering
  \begin{minipage}[b]{7.2 cm}
    \includegraphics[width=7.2cm]{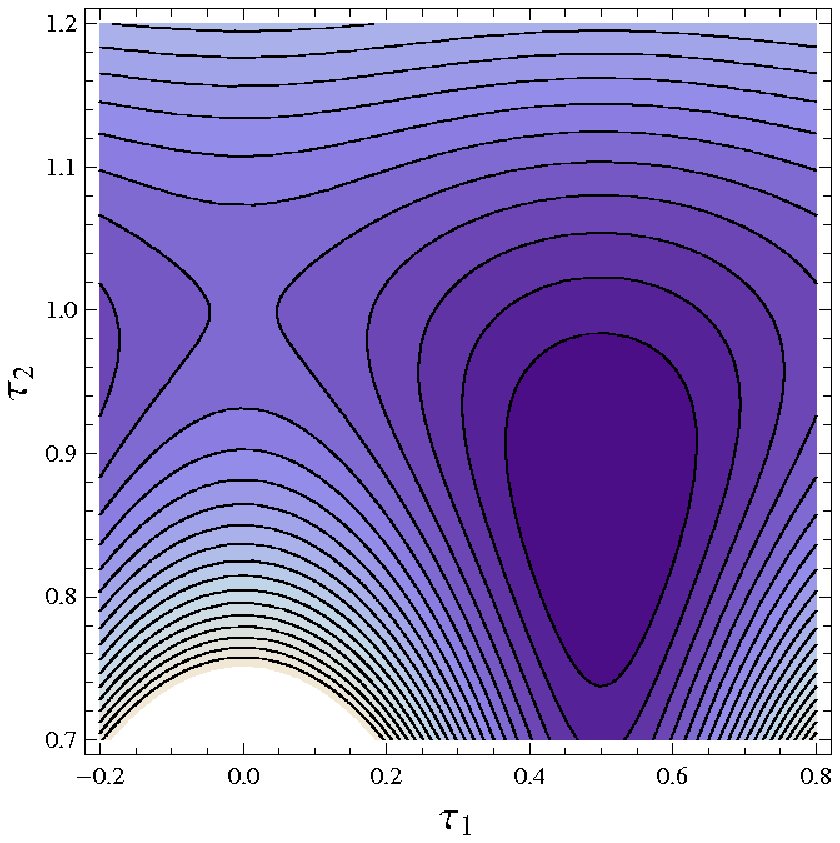}  
  \end{minipage}\hspace*{1cm}
  \begin{minipage}[b]{7.2 cm}
    \includegraphics[width=7.2cm]{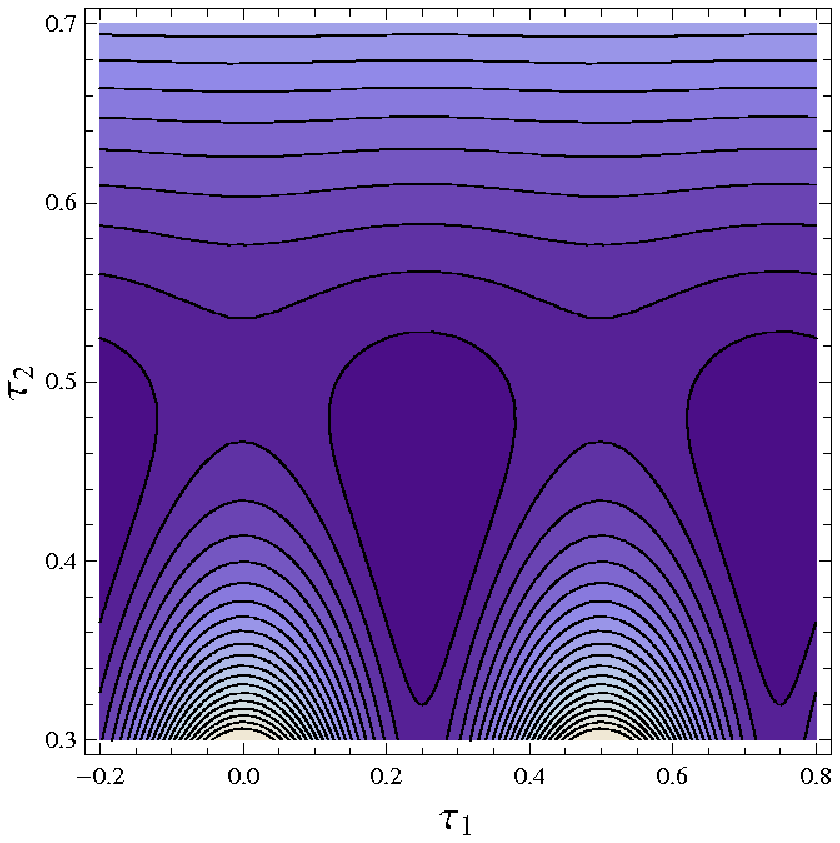}  
  \end{minipage}\vspace*{1cm}
  \begin{minipage}[b]{7.2 cm}
    \includegraphics[width=7.2cm,height=6cm]{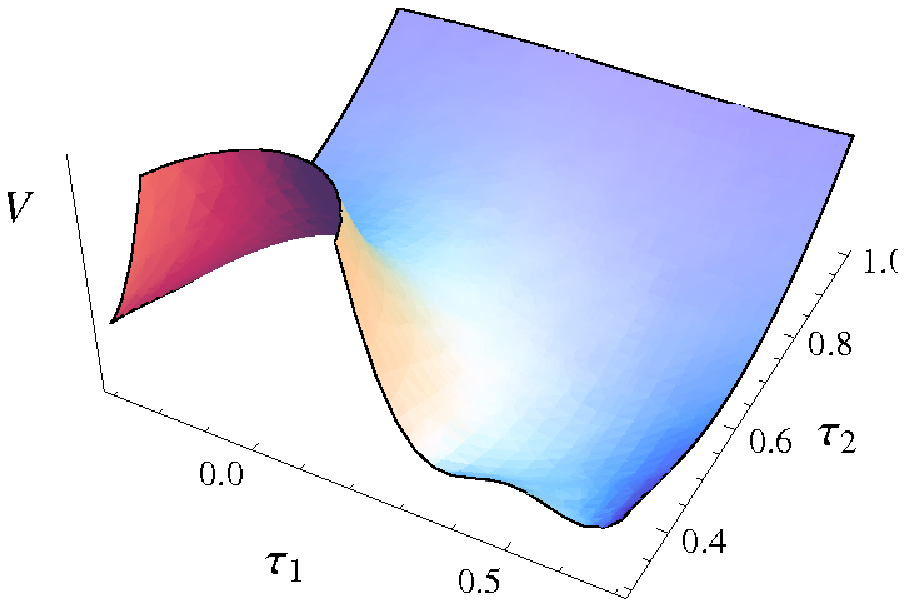}  
  \end{minipage}\hspace*{1cm}
  \begin{minipage}[b]{7.2 cm}
    \includegraphics[width=7.2cm]{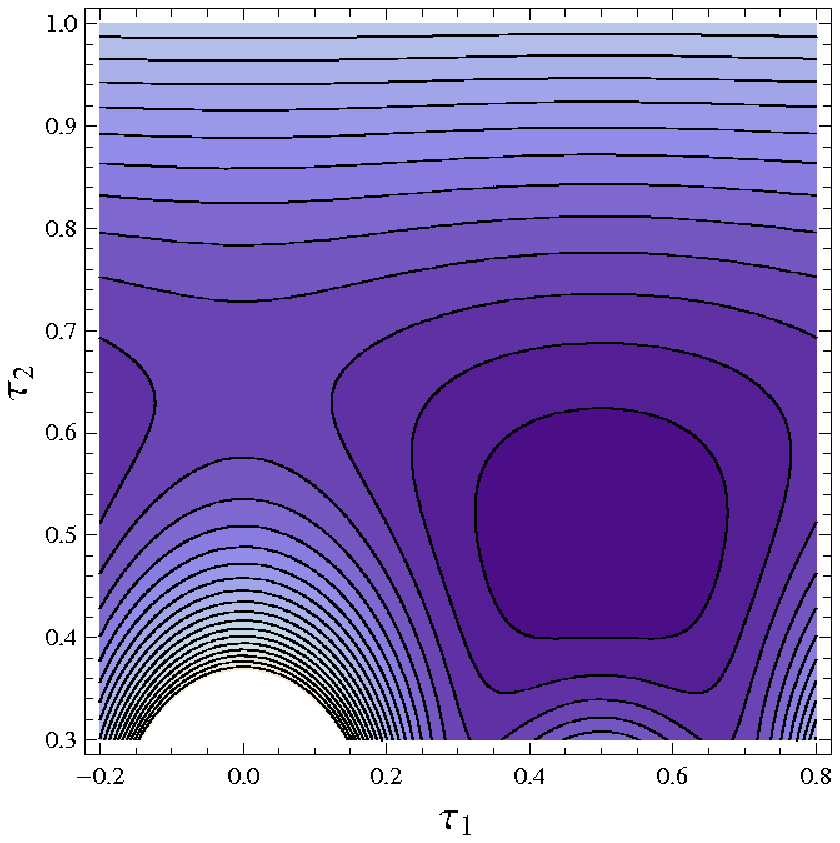}  
  \end{minipage}
  \caption{Effective potential for the shape moduli $\tau_1$ and $\tau_2$.
In the upper left (right) panel we plot the potential for $V_M^{0,0}$ 
($V_M^{0,0}$+$V_M^{0,1/2}$). In the lower panels we show the potential for 
the shape moduli in the given model. Note that the scaling in the $\tau_2$ 
direction is different. The different periodicities of the potential 
in the  $\tau_1$ direction correspond to different values of the parameter b 
in the modular transformations. For the given model there is a local minimum 
of the full potential at $(\tau_1,\tau_2) = (1/2,1/2)$ and a saddle point at 
$(\tau_1,\tau_2) = (0,1/\sqrt{2})$.} 
  \label{fig:potential}
\end{figure}
%
%
%
\begin{table}[t]
  \centering
  \renewcommand{\arraystretch}{1.7}
  \begin{tabular}{||l||c|c|c|c||c||}
    \hline
 & a& b & c & d & Fixed Points ($\tau_1,\tau_2$) \\
    \hline 
 $V^{0,0}$ & 0 mod 1 & 0 mod 1 & 0 mod 1 & 0 mod 1 & $(0,1)$ , $(1/2,\sqrt{3}/2)$ \\
    \hline  
 $V^{0,1/2}$ & 0 mod 1 & 0 mod 1 & 0 mod 2 & 1 mod 2 &  $(1/2,1/2)$ \\
    \hline  
$V^{1/2,0}$ & 1 mod 2 & 0 mod 2 & 0 mod 1 & 0 mod 1 &  $(1,1)$ \\
    \hline 
$V^{1/2,1/2}$ & 1 mod 2 & 0 mod 2 & 0 mod 2 & 1 mod 2 &   \\
              & 0 mod 2 & 1 mod 2 & 1 mod 2 & 0 mod 2 &  $(0,1)$ \\
    \hline 
$\Gamma(2)$ & 1 mod 2 & 0 mod 2 & 0 mod 2 & 1 mod 2 &  -- \\
    \hline 
 $V^{0,0}$+$V^{0,1/2}$ & 0 mod 1 & 0 mod 1/2 & 0 mod 2 & 0 mod 1 &  $(0,1/2)$ , $(1/4,\sqrt{3}/4)$  \\
    \hline  
$V_\text{casimir}$ & 0 mod 1 & 0 mod 1 & 0 mod 2 & 1 mod 2 &  $(1/2,1/2)$  \\
    \hline 
  \end{tabular}
  \caption{Modular symmetries (cf.~(\ref{modsym})) of different contributions 
           to the Casimir energy and the fixed points under those symmetries. 
           For general coefficients $A,B,C,D$ in (\ref{susy}), the symmetry corresponds to the 
           largest common subgroup, which is known as $\Gamma(2)$. 
           However, if the coefficients fulfill certain relations, the modular symmetry can
           be enhanced as shown in the last two lines.
    \label{tab:modsym}}    
\end{table} 
They are obtained by requiring
invariance of the Kaluza-Klein sums under the corresponding modular 
transformation, as discussed in Section~2. 
Naively, one would expect that adding two different contributions with 
different symmetries would lead to the largest common subgroup, which is 
given by $\Gamma(2)$. 
However, for certain relations between the coefficients $A$,$B$,$C$,$D$
there can be non-trivial cancellations, which lead to a larger modular symmetry.
For example, if the field content is such that $A=B$ and $D=C=0$, the
parameters of the modular group are restricted to $b=0$ mod $1/2$ and 
$c=0$ mod $2$. Surprisingly, the resulting symmetry is not only larger
than the symmetry of $V^{0,1/2}$, it is not even a subgroup of 
$\mathrm{SL}(2,\mathbbm{Z})$.

Fixed points under the modular symmetry are extrema of the effective 
potential, assuming that the volume is stabilized. Hence, minima of the 
effective potential may correspond to such fixed points. For fields with 
boundary condition $(+,+)$, this is indeed the case. The Casimir energy 
then has a minimum at $(\tau_1,\tau_2)=(1/2,\sqrt{3}/2)$ and a saddle point at
$(\tau_1,\tau_2)=(0,1)$ \cite{pp01}.
This implies that the shape moduli are stabilized at a torus lattice with 
$R_1=R_2$ and $\theta=\pi / 3$, which corresponds to the root lattice of 
the Lie algebras $\mathrm{SU}(3)$ or $\mathrm{G}_2$. 
For our example with $A=B$ and $D=C=0$ on the other hand, 
there is a minimum at  $(\tau_1,\tau_2)=(1/4,\sqrt{3}/4)$
and a saddle point at  $(\tau_1,\tau_2)=(0,1/2)$.
The minimum corresponds to the lattice with $R_1=2 R_2$ and $\theta=\pi / 3$.

\newpage

Let us now turn to our model. The wanted repulsive behavior of the Casimir
energy at small volume can be obtained if the contribution of particular bulk
hypermultiplets dominates \cite{bcs08},
\begin{eqnarray}\label{casimir}
V_\text{casimir} &=&
12 \left(V_{m_H}^{0,0}- V^{0,0}\right)
+ 12 \left(V_{m_H}^{0,1/2}- V^{0,1/2}\right) \nonumber \\
&&+ 8 \left(V_{m_H}^{1/2,0}- V^{1/2,0}\right)
+8 \left(V_{m_H}^{1/2,1/2}- V^{1/2,1/2}\right) \ ,
\end{eqnarray}
with $m_H^2 = -\lambda'\mu^2/(\mathcal{A}L^2\Lambda^2)$ and
$\lambda' < 0$, $|\lambda'| > |\lambda|, |\lambda''|$ (cf.~(\ref{masses})).
Remarkably, the potential has an enhanced modular symmetry compared to
$\Gamma(2)$. The allowed transformations have $c=0 \mod 2$ and 
$d=1 \mod 2$, with $a$ and $b$ $\in \mathbbm{Z}$.

Solving the fixed point equation (\ref{eq:fp}),
one finds a fixed point in the upper half-plane: 
$(\tau_1,\tau_2) = (1/2,1/2)$ with $a=-b=-d=1$ and $c=2$. It
corresponds to a minimum in the effective potential.
There is also a saddle point at $(\tau_1,\tau_2) = (0,1/\sqrt{2})$.
For the minimum, the torus lattice again has an enhanced symmetry: $R_1=\sqrt{2}R_2$
and $\theta=\pi /4$, which corresponds to the root lattice of SO(5).
Its discrete symmetry is $\mathbbm{Z}_4$.

\subsection{Volume Modulus}

In \cite{bcs08} it has been shown that spontaneous gauge symmetry breaking by
bulk Higgs fields together with supersymmetry breaking can stabilize the compact
dimensions at the GUT scale. The detailed mechanism of supersymmetry breaking is
discussed in \cite{bms09}.
Consider the breaking of $U(1)_X$ as discussed in 
Section~\ref{model}. In orbifold compactifications of the heterotic string 
a vacuum expectation value $\langle \Phi^c\rangle$ can be induced by localized 
Fayet-Iliopoulos (FI) terms. Vanishing of the $D$-terms then implies
\begin{align}
\langle\Phi^c\rangle^2=\frac{C\Lambda^2}{\mathcal{A}L^2}\ ,
\end{align}
where $C \ll 1$ is a loop factor and $\Lambda$ is the string scale or, more
generally, the UV cutoff of the model. The expectation value is volume
suppressed because $\Phi^c$ is a bulk field and the FI-terms are localized
at fixed points. In terms of the bulk Higgs mass (\ref{higgsmass}) one obtains, 
using $g_6^2/(\mathcal{A}L^2) = g_4^2 \simeq 1/2$,
\begin{equation}
\mathcal{M}^2 \simeq g_6^2 \langle\Phi^c\rangle^2
\simeq \frac{1}{2} C \Lambda^2 \ .
\end{equation}
For orbifold compactifications of the heterotic string one finds 
$\mathcal{M} \sim M_{\mathrm{GUT}}$.

Supersymmetry breaking by a brane field
$S$, with $\mu = F_S/\Lambda$, leads to a `classical' contribution to the 
vacuum energy density,
\begin{align}
V_{\mathrm{cl}}
&= \int \text{d}^2y \int \text{d}^4\theta \sqrt{\widetilde{g}} \, \delta^2(y) 
\left\langle S^\dagger S \left( 1-\frac{\lambda''}{\Lambda^4}
(\Phi^\dagger\Phi +\Phi^{c\dagger}\Phi^c)\right)\right\rangle \nonumber\\
&= \frac{F_S^2}{\mathcal{A}^2} - 
\frac{2\lambda''\mu^2\mathcal{M}^2}{\mathcal{A}^3 L^2\Lambda^2}
+ \ldots \ ,\label{classical}
\end{align} 
where the first term is a tree level potential \cite{bms09},
and $\tilde{g}_{\mu\nu} = \mathcal{A}^{-1}\eta_{\mu\nu}$ is the induced 
metric
at the fixed point $y=0$.
The first term proportional to $F_S^2$ will be absorbed into the brane 
tension.

\begin{figure}
  \centering
\psfrag{A}{$\mathcal{A}$}
    \includegraphics[width=8.8cm]{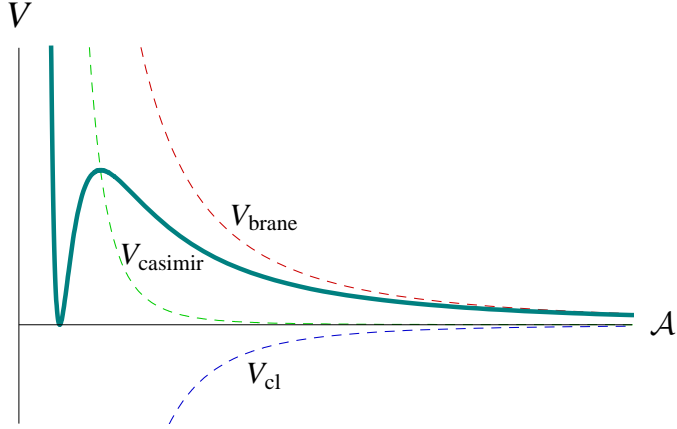}  
\caption{Effective potential for the volume modulus (full line). The different 
contributions to the potential are also shown separately; $V_\text{brane}$ 
represents the brane counterterm.}
\label{fig:potential}
\end{figure}

In the vicinity of $\tau_1=1/2$, the Casimir energy is to a good approximation 
given by 
\begin{align} 
V^{\alpha,\beta}_M 
=&+\frac{M^6 L^2}{3072\pi^3 \mathcal{A}} \left[\frac{11}{12} 
-\log\left(\frac{M}{\sqrt{\mathcal{A}}\mu_r}\right)\right]
\nonumber \\
&- \frac{M^4}{64\pi^2\mathcal{A}^2 }
\left[\frac{3}{4}-\log\left(\frac{M}{\sqrt{\mathcal{A}}\mu_r}\right)\right]
\delta_{\alpha 0}\delta_{\beta 0}
\nonumber \\
&- \frac{M^3 \tau_2^{3/2}}{4 \pi^3\mathcal{A}^{5/2}L} 
\sum_{p=1}^{\infty}\frac{\cos(2\pi p \alpha)}{p^3}
K_{3}\left(p \tfrac{\sqrt{\mathcal{A}}L M}{2\sqrt{\tau_2}}\right) 
 \nonumber \\
&- \frac{M^3 \tau_2^{-3/2}}{32 \pi^3\mathcal{A}^{5/2}L} 
\sum_{p=1}^{\infty}\frac{\cos(2\pi p \alpha)}{p^3}
K_{3}\left(p \sqrt{\mathcal{A} \tau_2}L M\right) 
 \nonumber \\
&+ \frac{M^4}{64 \pi^3 \tau_2^3\mathcal{A}^2}
\sum_{p=1}^{\infty}\frac{\cos(2\pi p \alpha)}{p^2}
K_{4}\left(p \sqrt{\mathcal{A} \tau_2}L M \right)\left(\tau_1-\tfrac{1}{2}
\right)^2\ ,
\end{align}
where we have performed a Taylor expansion around $\tau_1=1/2$.
This approximation is valid for small $\tau_2$, and we have dropped additional 
terms in $V^{0,\beta}_M$, which cancel in the sum of $V^{0,0}_M$ and 
$V^{0,1/2}_M$.

Expanding the Bessel functions for small arguments and performing the
summations over $p$, we obtain for the four different contributions
\begin{align}
V_M^{0,0}(\mathcal{A},\tau_1,\tau_2) =& 
-\frac{16\pi^3\tau_2^3}{945\mathcal{A}^4 L^4}
-\frac{\pi^3}{3780\mathcal{A}^4 L^4\tau_2^3}
+\frac{\pi^3(\tau_1-1/2)^2}{1260\mathcal{A}^4 L^4\tau_2^5} \nonumber \\
&+ \frac{\pi M^2\tau_2^2}{180\mathcal{A}^3L^2}
+ \frac{\pi M^2}{2880\mathcal{A}^3L^2\tau_2^2}
- \frac{\pi M^2(\tau_1-1/2)^2}{5760\mathcal{A}^3L^2\tau_2^2}\ , \label{exp1} \\
V_M^{0,1/2}(\mathcal{A},\tau_1,\tau_2) =& 
-\frac{16\pi^3\tau_2^3}{945\mathcal{A}^4 L^4}
-\frac{\pi^3}{3780\mathcal{A}^4 L^4\tau_2^3}
+\frac{\pi^3(\tau_1-1/2)^2}{1260\mathcal{A}^4 L^4\tau_2^5} \nonumber \\
&+ \frac{\pi M^2\tau_2^2}{180\mathcal{A}^3L^2}
+ \frac{\pi M^2}{2880\mathcal{A}^3L^2\tau_2^2}
- \frac{\pi M^2(\tau_1-1/2)^2}{5760\mathcal{A}^3L^2\tau_2^2}\ , \label{exp2} \\
V_M^{1/2,0}(\mathcal{A},\tau_1,\tau_2) =& 
+\frac{31\pi^3\tau_2^3}{1890\mathcal{A}^4 L^4}
+\frac{31\pi^3}{120960\mathcal{A}^4 L^4\tau_2^3}
-\frac{31\pi^3(\tau_1-1/2)^2}{40320\mathcal{A}^4 L^4\tau_2^5} \nonumber \\
&- \frac{7\pi M^2\tau_2^2}{1440\mathcal{A}^3L^2}
- \frac{7\pi M^2}{23040\mathcal{A}^3L^2\tau_2^2}
+ \frac{7\pi M^2(\tau_1-1/2)^2}{46080\mathcal{A}^3L^2\tau_2^2}\ ,\label{exp3}\\
V_M^{1/2,1/2}(\mathcal{A},\tau_1,\tau_2) =& 
+\frac{31\pi^3\tau_2^3}{1890\mathcal{A}^4 L^4}
+\frac{31\pi^3}{120960\mathcal{A}^4 L^4\tau_2^3}
-\frac{31\pi^3(\tau_1-1/2)^2}{40320\mathcal{A}^4 L^4\tau_2^5} \nonumber \\
&- \frac{7\pi M^2\tau_2^2}{1440\mathcal{A}^3L^2}
- \frac{7\pi M^2}{23040\mathcal{A}^3L^2\tau_2^2}
+ \frac{7\pi M^2(\tau_1-1/2)^2}{46080\mathcal{A}^3L^2\tau_2^2}\ . \label{exp4}
\end{align}
The total effective potential is now given by the sum of the Casimir energy
(\ref{casimir}), the classical energy density (\ref{classical}) and a brane 
tension,
\begin{equation}\label{vtot0}
V_{\text{tot}}(\mathcal{A},\tau_1,\tau_2) = 
V_{\text{casimir}}(\mathcal{A},\tau_1,\tau_2) + 
V_{\text{cl}}(\mathcal{A}) + V_{\text{brane}}(\mathcal{A})\ .
\end{equation} 
Inserting the expansions (\ref{exp1})-(\ref{exp4}) into the expression for
the Casimir energy, one finally obtains
\begin{equation}\label{vtot}
V_{\text{tot}}(\mathcal{A},\tau_1,\tau_2)
= -\frac{\pi\lambda'\mu^2}{288 \mathcal{A}^4 L^4\Lambda^2}
\left(16\tau_2^2+\tau_2^{-2}-\frac{2 (\tau_1-1/2)^2}{\tau_2^{4}}\right) 
- \frac{2\lambda''\mu^2 \mathcal{M}^2}{\mathcal{A}^3 L^2\Lambda^2} 
+\frac{\kappa}{\mathcal{A}^2}\ , 
\end{equation}
where
\begin{equation}
\kappa = -\frac{36\lambda''^2\mu^2\mathcal{M}^4}{\pi\lambda'\Lambda^2} > 0\ .
\end{equation} 
The brane tension $\kappa$ has been adjusted such that the potential
$V_{\text{tot}}$ vanishes at the local minimum. The different contributions to 
the effective potential are shown in Figure~4.

As discussed in the previous section, the Casimir energy, and therefore
$V_{\text{tot}}$, has a local minimum at $\tau_1 = \tau_2 = 1/2$.
The volume modulus is then fixed at
\begin{align}\label{a0}
\mathcal{A}_0 L^2 
=-\frac{\pi\lambda'}{36\lambda''}\frac{1}{\mathcal{M}^2} \ .
\end{align}
For $|\lambda'| > \lambda''$, as required by a repulsive Casimir energy
at small volume, one then obtains stabilization of the compact dimensions
at the inverse GUT scale, 
$\sqrt{\mathcal{A}L^2} \sim 1/\mathcal{M} \sim 1/M_{\mathrm{GUT}}$.

\subsection{Moduli Masses}

The moduli fields $\mathcal{A}, \tau_{1}$  and $\tau_{2}$ have masses much
smaller than the inverse size of the compact dimensions. Their Lagrangian is
obtained by dimensional reduction (cf.~\cite{pp01}) and from the effective
potential (\ref{vtot0}),
\begin{equation}
\mathcal{L} = \sqrt{g} \left\{\frac{M^{4}_6 L^2}{2} \left( R(g) 
+ \frac{g^{\mu\nu}\partial_{\mu} \mathcal{A}
\partial_{\nu}\mathcal{A}}{\mathcal{A}^2} + \frac{g^{\mu\nu} 
\partial_{\mu}\tau\partial_{\nu} \tau^*}{2\mathcal{\tau}_2^2} \right) 
-  V_{\textrm{tot}}(\mathcal{A},\tau_{1},\tau_{2}) \right\} \ .
\label{action}
\end{equation}
After a constant Weyl rescaling, $g_{\mu\nu} =\mathcal{A}_0 \bar{g}_{\mu\nu}$
(cf.~(\ref{a0})), the Lagrangian for the moduli depends on 
$\mathcal{A}_0$ and the 4D Planck mass $M_4 = \sqrt{\mathcal{A}_0 L^2}M_6^2$,
\beq
\mathcal{L_M} = \sqrt{\bar{g}} 
\left\{ \frac{M^{2}_4}{2} \left( \frac{\bar{g}^{\mu\nu} \partial_{\mu} 
\mathcal{A}  \partial_{\nu} \mathcal{A}}{\mathcal{A}^2} +
\frac{\bar{g}^{\mu\nu}
\partial_{\mu}\tau \partial_{\nu}\tau^*}{2\mathcal{\tau}_2^2} \right)
- \mathcal{A}_0^2\ V_{\textrm{tot}}(\mathcal{A},\tau_{1},\tau_{2}) \right\}\ .
\label{lagrangian}
\eeq
Expanding the moduli fields around the minimum,
\begin{equation}
\mathcal{A} = \mathcal{A}_0 + \frac{\mathcal{A}_0}{M_4}\ \bar{\mathcal{A}}\ , 
\quad
\tau_{1,2} = \frac{1}{2} + \frac{1}{\sqrt{2}M_4}\ \bar{\tau}_{1,2}\ ,
\end{equation}
yields the Lagrangian for the canonically normalized fluctuations,
\begin{eqnarray}
\mathcal{L_M} &=& \sqrt{\bar{g}} 
\Bigg\{\frac{1}{2}\left(\bar{g}^{\mu\nu} \partial_{\mu} \bar{\mathcal{A}}  
\partial_{\nu} \bar{\mathcal{A}} +
\bar{g}^{\mu\nu}\partial_{\mu}\bar{\tau}_1\partial_{\nu}\bar{\tau}_1 + 
\bar{g}^{\mu\nu}\partial_{\mu}\bar{\tau}_2\partial_{\nu}\bar{\tau}_2\right) 
\nonumber\\
&&   \hspace{1cm}
-\frac{\mathcal{A}_0^2}{M^{2}_4}\left(\frac{\mathcal{A}^2_0}{2}
\frac{\widehat{\partial^2 V_{\textrm{tot}}}}{\partial \mathcal{A}^2} 
\bar{\mathcal{A}}^2 
+ \frac{1}{4}\frac{\widehat{\partial^2 V_{\textrm{tot}}}}{\partial\tau_1^2}
\bar{\tau}_1^2
+ \frac{1}{4}\frac{\widehat{\partial^2 V_{\textrm{tot}}}}{\partial\tau_2^2} 
\bar{\tau}_2^2 \right) + \ldots \Bigg\}\ , 
\end{eqnarray}
where the hat denotes that the second derivatives of $V_{\textrm{tot}}$ are
evaluated at the minimum.
Together with Eqs.~(\ref{vtot}) and (\ref{a0}) we now obtain the moduli masses
\begin{align}
m^{2}_{\mathcal{A}} &=  \frac{\mathcal{A}_0^4}{M^{2}_4} 
\frac{\widehat{\partial^2 V_{\textrm{tot}}}}{\partial \mathcal{A}^2} 
= \frac{\lambda^{\prime\prime}}{\mathcal{A}_0 L^2}
\frac{2 \mathcal{M}^2 \mu^2}{\Lambda^2 M^2_4}\ ,\\
m^{2}_{\tau_2} &= \frac{\mathcal{A}_{0}^2}{2 M^{2}_4}
\frac{\widehat{\partial^2 V_{\textrm{tot}}}}{\partial \tau_2^2} 
= 4 m^{2}_{\mathcal{A}} \ ,\\
m^{2}_{\tau_1} &\simeq m^{2}_{\tau_2}\ , 
\end{align}
which depend on the scale of supersymmetry breaking $\mu$, the cutoff $\Lambda$
and the size of the compact dimensions 
$\sqrt{\mathcal{A}_0 L^2} \sim 1/\mathcal{M} > 1/\Lambda$.
The mass $m_{\tau_1}$ has been obtained numerically, based on the complete
expression (\ref{casimir}) for $V_{\textrm{casimir}}$, since the analytical 
result (\ref{vtot}) away from $\tau_1=1/2$ only holds for small $\tau_2$ and not at the minimum 
$\tau_1 = \tau_2 = 1/2$. 

The moduli masses can be related to the gravitino mass using 
$\mu = F_S/\Lambda$ and $m_{3/2}= F_S/(\sqrt{3}M_4)$, which yields 
\begin{equation}
m^{2}_{\mathcal{A}} = 
\frac{6\lambda^{\prime\prime}\mathcal{M}^2}{\mathcal{A}_0 L^2\Lambda^4} 
m_{3/2}^2\ .
\end{equation}
For a compactification scale 
$\sqrt{\mathcal{A}_0 L^2} \sim 1/\mathcal{M}$, one obtains
\begin{equation}\label{volmass}
m^{2}_{\mathcal{A}} = 
\frac{6\lambda^{\prime\prime}}{\mathcal{A}_0^2 L^4 \Lambda^4} m_{3/2}^2\ ,
\end{equation}
i.e., the moduli masses are volume suppressed compared to the gravitino
mass \cite{bms09}.
 
An upper bound on the coupling $\lambda^{\prime\prime}$ of the brane field 
$S(x)$ to the bulk field $\Phi(x,y)$, and therefore on the moduli masses, can
be obtained by naive dimensional analysis (NDA) \cite{Chacko:1999hg}.
For this purpose, one rewrites the relevant part of the 6D Lagrangian
\begin{equation} \label{eq:LDCanonical}
	\mathscr{L} =
	\hat{\mathscr{L}}_\mathrm{bulk}(\Phi(x,y)) +
	\delta^{2}(y-y_S) \, \hat{\mathscr{L}}_{S}(\Phi(x,y),S(x))
\end{equation}
in terms of dimensionless fields $\hat \Phi(x,y)$ and $\hat S(x)$,
and the cutoff $\Lambda$,
\begin{equation} \label{eq:LDDimless}
	\mathscr{L} =
	\frac{\Lambda^6}{\ell_6/C} \,
	 \mathscr{\hat L}_\mathrm{bulk}(\hat \Phi(x,y)) +
	 \delta^{2}(y-y_S) \, \frac{\Lambda^4}{\ell_4/C} \,
	 \mathscr{\hat L}_{S}(\hat \Phi(x,y),\hat S(x)) \;,
\end{equation}
where $\ell_6 = 128 \pi^3$ and $\ell_4 = 16\pi^2$; the factor $C$ accounts
for the multiplicity of fields in loop diagrams, with $C=8$ in the present
model (cf.~\cite{bks05}). The rescaling of chiral bulk and brane superfields 
reads  
\begin{equation}
\Phi(x,y) = \frac{\Lambda}{\sqrt{\ell_6/C}}\hat{\Phi}(x,y)
\quad , \quad
S(x) = \frac{\Lambda}{\sqrt{\ell_4/C}}  \hat S(x) \;.
\label{eq:phiAndphiHat}
\end{equation}
The ratio $C/\ell_D$ gives the typical suppression of loop diagrams.  This 
suppression is canceled by the factors $\ell_6/C$ and $\ell_4/C$ in front of 
the Lagrangians $\mathscr{\hat L}$ in
Eq.~\eqref{eq:LDDimless}.  Consequently, all loops will be of the same order
of magnitude, provided that all couplings are $\mathcal{O}(1)$.  Thus,
according to the NDA recipe the effective 6D theory remains
weakly coupled up to the cutoff $\Lambda$, if the dimensionless
couplings in Eq.~\eqref{eq:LDDimless} are smaller than one. 

Let us now apply the NDA recipe to the coupling $\lambda^{\prime\prime}$. 
Using Eq.~\eqref{eq:phiAndphiHat}, we obtain
\begin{equation}
\mathscr{L}_S \supset \frac{\Lambda^4}{\ell_4/C} \ \int 
\frac{\text{d}^4 \theta}{\Lambda^2}\ \frac{\lambda^{\prime\prime}C}{\ell_6}  
\hat{S}^\dagger \hat{S}
\left(\hat{\Phi}^\dagger\hat{\Phi} +\hat{\Phi}^{c\dagger}\hat{\Phi}^c\right)\ .
\end{equation}
The NDA requirement that all couplings be smaller than one implies 
$\lambda^{\prime\prime} \lesssim \ell_6/C = 16 \pi^3$. 
This translates into
\begin{equation}
m^{2}_{\mathcal{A}} \lesssim \frac{96\pi^3}{\mathcal{A}_0^2 L^4 \Lambda^4} 
m_{3/2}^2\ .
\end{equation}
However, this bound cannot be saturated, since the same bound
holds for $|\lambda'| > \lambda^{\prime\prime}$. 
Further, one has $\Lambda \simeq M_6 \simeq 10\ M_{\mathrm{GUT}}$ in the model under 
consideration \cite{bks05}. This, together with the bound on $\lambda^{\prime\prime}$,
leads to the estimate
\begin{equation}\label{modulibound}
m^{2}_{\mathcal{A}} \lesssim  0.1\ m_{3/2}^2\ .
\end{equation}
Hence, all moduli masses are smaller than the gravitino mass.

It is instructive to compare the upper bound on the moduli masses with the 
upper bound on the gaugino mass (\ref{masses}), 
\begin{equation}
m_g = \frac{h\mu}{\mathcal{A}L^2\Lambda^2} 
\simeq \frac{\sqrt{3}h}{\sqrt{\mathcal{A}_0L^2}\Lambda}m_{3/2}\ .
\end{equation}
Compared to the moduli masses (\ref{volmass}), the gaugino
mass is weaker volume suppressed. Correspondingly, the NDA analysis allows
the gaugino mass to be larger or smaller than the gravitino mass 
\cite{bhk05}.

\section{Conclusions}

We have studied a 6-dimensional orbifold GUT model, compactified on a
$T^2/\mathbbm{Z}_2$ orbifold with two Wilson lines. The Casimir energy
depends on the boundary conditions of the various bulk fields and is
a function of the shape moduli. It is remarkable that the
minimum of the effective potential occurs at a point in field space
where the torus lattice has an `enhanced symmetry' corresponding to
the root lattice of $\mathrm{SO(5)}$. 

The $\mathrm{SO(5)}$ lattice has a discrete $\mathbbm{Z}_4$ symmetry
which is larger than the $\mathbbm{Z}_2$ symmetry of a generic torus. 
Vacua with unbroken discrete symmetries are phenomenologically desirable
since they can explain certain features of the supersymmetric standard 
model, in particular the difference between Higgs and matter fields.
Our analysis suggests that such discrete symmetries may arise dynamically
in the compactification of higher-dimensional field and string theories.

The interplay of a repulsive Casimir force at small volume and an attractive
interaction generated by the coupling of a bulk Higgs field to a supersymmetry
breaking brane field stabilizes the volume modulus at the GUT scale, which
is determined by the size of localized Fayet-Iliopoulos terms. The masses
of shape and volume moduli are smaller than the gravitino mass.

A full supergravity treatment of the described
stabilization mechanism still remains to be worked out. 
Also the phenomenological and cosmological consequences of moduli fields
lighter than the gravitino require further investigations.

\section*{Acknowledgments}
We would like to thank
A.~Hebecker, M.~Klaput, O.~Lebedev, J.~Louis, J.~M\"oller, C.~Paleani, 
M.~Ratz, J.~Schmidt and J.~Teschner for helpful discussions.
This work has been supported by the SFB-Transregio 27 ``Neutrinos and 
Beyond'' and by the DFG cluster of excellence ``Origin and 
Structure of the Universe''.

\newpage
\appendix

\section{Evaluation of Casimir Sums}

Our evaluation of the Casimir double sums requires two single sums which
we shall now consider. 
The first sum reads
\begin{align}
 \widetilde{F}(s;a,c) 
\equiv \sum_{m=0}^{\infty}\frac{1}{\left[(m+a)^2+c^2\right]^{s}}\;.
\end{align}
This is a series of the generalized Epstein-Hurwitz zeta type. 
The result can be found in \cite{eli94} 
and is given by
\begin{align}
 \widetilde{F}(s;a,c) =& \frac{c^{-2s}}{\Gamma(s)}
\sum_{m=0}^{\infty}\frac{(-1)^m\Gamma(m+s)}{m!}c^{-2m}\zeta_H(-2m,a)
+ \sqrt{\pi}\frac{\Gamma(s-\tfrac{1}{2})}{2\Gamma(s)}c^{1-2s}\nonumber \\
&+\frac{2\pi^s}{\Gamma(s)}c^{1/2-s}\sum_{p=1}^{\infty}p^{s-1/2}\cos(2\pi p a)
K_{s-1/2}(2\pi p c) \;,
\end{align}
where $\zeta_H(s,a)$ is the Hurwitz zeta-function.
Note that this is not a convergent series but an asymptotic one.
In the following it will be important that
$\zeta_H(-2m,0)=\zeta_H(-2n,1/2)=0$ for $m\in \mathbbm{N}$ and $n\in 
\mathbbm{N}_0$. In our case, the first sum in $\widetilde{F}(s;a,c)$ thus 
reduces to a single term. For $a=1/2$ the sum vanishes, and for $a=0$ only
the first term contributes; with $\zeta_H(0,0)=1/2$ one obtains $c^{-2s}/2$.

The second, related sum is given by
\begin{align}
\label{SingleSum}
F(s;a,c) \equiv 
\sum_{m=-\infty}^\infty\frac{1}{\left[(m + a)^{2} + c^{2}\right]^s}\;. 
\end{align}
Using the two identities ($m\in \mathbbm{N}$)
\begin{align}
\zeta_H(-2m,a) &=-\zeta_H(-2m,1-a) \;, \\ 
F(s;a,c) &= \widetilde{F}(s;a,c) + \widetilde{F}(s;1-a,c)\;,
\end{align}
one easily obtains, in agreement with \cite{pp01},
\begin{equation}
F(s;a,c) = \frac{\sqrt{\pi}}{\Gamma(s)} |c|^{1-2s} 
\left[\Gamma\left(s-\tfrac{1}{2}\right)
+ 4\sum_{p=1}^{\infty} \cos(2\pi p a)(\pi\,p\,|c|)^{s -\frac{1}{2}} 
K_{s -\frac{1}{2}} (2\pi \,p \,|c|) \right] \,.
\end{equation}
These two sums provide the basis for our evaluation of the Casimir sums.

\subsection[Casimir Sum (I) on $T^2/{\mathbbm Z}_2^3$] 
{Casimir Sum (I) on ${ \bf T^2/\mathbbm Z}_2^3$} 

We first consider the summation
\begin{align}
 \left[\sum \right]_{m,n} = 
\sum_{m=0}^{\infty}\sum_{n=-\infty}^{\infty} \;. 
\end{align}
In this case the Casimir energy is obtained from
\begin{align}
\sum_{m=0}^{\infty}\sum_{n=-\infty}^{\infty}
\left[(n+\beta-(m+\alpha)\tau_1)^2 + (m+\alpha)^2 \tau_2^2  +  \kappa^2\right]^{-s}   \;. 
\end{align}
where we have shifted $s\to s+2$ and defined 
$\kappa^2=  \frac{\mathcal{A}L^2\tau_2}{4(2\pi)^2}   M^2$.
Using the expression for $F(s;a,c)$ we can perform the sum over $n$,
\begin{align}
&\sum_{m=0}^{\infty}\sum_{n=-\infty}^{\infty}
\left[(n+\beta-(m+\alpha)\tau_1)^2 + (m+\alpha)^2 \tau_2^2  +  \kappa^2\right]^{-s} 
  \nonumber \\
=& \;
\sqrt{\pi} \frac{ \Gamma(s-\frac{1}{2}) }{\Gamma(s)} 
 \sum_{m=0}^{\infty}(\tau_2^2(m+\alpha)^2 +\kappa^2)^{1/2-s} 
 \nonumber \\
& \; +  \frac{4\sqrt{\pi}}{\Gamma(s)} \sum_{p=1}^{\infty} 
\cos(2 \pi p(\beta-(m+\alpha)\tau_1) ) 
 \sum_{m=0}^{\infty}(\pi \,p)^{s-\tfrac{1}{2}}
\left(\sqrt{\tau_2^2(m+\alpha)^2 +\kappa^2}\right)^{\frac{1}{2}-s}  \nonumber \\
& \; \times \;
K_{s -\frac{1}{2}} \left(2\pi \,p \,\sqrt{\tau_2^2(m+\alpha)^2 +\kappa^2}\right) \nonumber \\
\equiv& f_1(s) + f_2(s) 
\;.
\end{align}
Let us consider $f_1(s)$ first. The sum over $m$ can be performed with the
help of $\widetilde{F}(s;a,c)$,
\begin{align}
f_1(s)
=& \;
\sqrt{\pi} \frac{ \Gamma(s-\frac{1}{2}) }{\Gamma(s)} 
 \sum_{m=0}^{\infty}(\tau_2^2(m+\alpha)^2 +\kappa^2)^{1/2-s} 
 \nonumber \\
=& \; \sqrt{\pi}\frac{\Gamma(s-1/2)}{\Gamma(s)} \kappa^{1-2s}
\zeta_H(0,\alpha)  +  \frac{\pi}{2(s-1)}
\frac{\kappa^{2-2s}}{\tau_2}\nonumber \\
& +\;\frac{2\pi^{s}}{\Gamma(s)} \tau_2^{-s}\kappa^{1-s}
\sum_{p=1}^{\infty}p^{s-1}\cos(2\pi p \alpha)
K_{s-1}\left(2\pi p \left(\tfrac{\kappa}{\tau_2}\right)\right) 
\end{align}
Recalling the shift in $s$,
we can now write $\zeta(s)$ as 
\begin{align}
\zeta(s)=& \frac{\mu_r^{2s+4} 4^{-s}(2\pi)^{-2s-4}\pi^2}
{2 \mathcal{A}^{-2s}L^{-2s}\tau_2^{-s} s(s+1)}
\bigg\{ \sqrt{\pi}\frac{\Gamma(s-1/2)}{\Gamma(s)} \kappa^{1-2s}
\zeta_H(0,\alpha)
\nonumber \\
&+ \frac{\pi}{2(s-1)}
\frac{\kappa^{2-2s}}{\tau_2}\nonumber \\
&+\frac{2\pi^{s}}{\Gamma(s)} \tau_2^{-s}\kappa^{1-s}
\sum_{p=1}^{\infty}p^{s-1}\cos(2\pi p \alpha)
K_{s-1}\left(2\pi p \left(\tfrac{\kappa}{\tau_2}\right)\right) 
 \nonumber \\
&+ \frac{4\sqrt{\pi}}{\Gamma(s)} \sum_{p=1}^{\infty} 
\cos(2 \pi p(\beta-(m+\alpha)\tau_1) ) 
 \sum_{m=0}^{\infty}(\pi \,p)^{s-\tfrac{1}{2}}
\left(\sqrt{\tau_2^2(m+\alpha)^2 +\kappa^2}\right)^{\frac{1}{2}-s}  \nonumber \\
& \hspace{2cm}
K_{s -\frac{1}{2}} \left(2\pi \,p \,\sqrt{\tau_2^2(m+\alpha)^2 +\kappa^2}\right)
\bigg\} \;.
\end{align}
Now we have to differentiate with respect to $s$ and set $s=-2$.
Since $\Gamma(-2)= \infty$, the derivative has only to act on $\Gamma(s)$
if the corresponding term is inversely proportional to $\Gamma(s)$.
Performing the differentiation, using
\begin{align}
\frac{\text{d}}{\text{d}s}\frac{1}{\Gamma(s)}\bigg|_{s=-2}
=-\frac{\Gamma^{'}(s)}{\Gamma(s)^2}\bigg|_{s=-2} = +2  \;,
\end{align}
as well as $K_{a}(z)=K_{-a}(z)$ and
substituting again $\kappa=\frac{\sqrt{\mathcal{A}\tau_2}M L}{4\pi}$ we finally obtain for the
Casimir energy,
\begin{align} 
V^{\alpha,\beta (I)}_M 
=& -\frac{4\pi^2}{\mathcal{A}^4L^4\tau_2^2} 
\bigg\{ -\frac{16\pi}{15}\frac{\mathcal{A}^{5/2}L^5\tau_2^{5/2}M^5}{(4\pi)^5} \zeta_H(0,\alpha)
\nonumber \\
&+ \frac{\pi\mathcal{A}^3 L^6 \tau_2^2M^6}{36 (4\pi)^6} \left[-11 
+12\log\left(\frac{M}{\sqrt{\mathcal{A}}\mu_r}\right)\right]
\nonumber \\
&+ \frac{4}{\pi^2} \tau_2^{2}\frac{\mathcal{A}^{3/2}L^3\tau_2^{3/2}M^3}{(4\pi)^3}
\sum_{p=1}^{\infty}\frac{\cos(2\pi p \alpha)}{p^3}
K_{3}(2\pi p \left(\tfrac{\sqrt{\mathcal{A}}L M}{4\pi\sqrt{\tau_2}}\right)) 
 \nonumber \\
&+  \frac{8}{\pi^2} \sum_{p=1}^{\infty} 
\frac{\cos(2 \pi p (\beta-(m+\alpha)\tau_1))}{p^{5/2}} 
 \sum_{m=0}^{\infty}
\left(\tau_2^2(m+\alpha)^2 + \frac{\mathcal{A}L^2\tau_2 M^2}{(4\pi)^2}  \right)^{\frac{5}{4}} \nonumber \\
& \quad K_{5/2} (2\pi \,p \,\sqrt{\tau_2^2(m+\alpha)^2 
+  \frac{\mathcal{A}L^2\tau_2 M^2}{(4\pi)^2}   })\bigg\} \;.
\end{align}

\subsection[Casimir Sum (II) on $T^2/{\mathbbm Z}_2^3$] 
{Casimir Sum (II) on ${ \bf T^2/\mathbbm Z}_2^3$} 

The second relevant summation is
\begin{align}
 \left[\sum \right]_{m,n} = 
 \left[ \delta_{0,m} \sum_{n=0}^{\infty}+
\sum_{m=1}^{\infty}\sum_{n=-\infty}^{\infty} \right] \;.
\end{align}
For the corresponding boundary conditions one has $\alpha=0$.
The Casimir sum can then be written as 
\begin{align}
& \left[ \delta_{0,m} \sum_{n=0}^{\infty}+
\sum_{m=0}^{\infty}\sum_{n=-\infty}^{\infty} 
-\delta_{m,0}\sum_{n=-\infty}^{\infty}   \right]
\left[(n+\beta-m\tau_1)^2 + m^2 \tau_2^2  +  \kappa^2\right]^{-s} \;,
\end{align}
where we again shifted $s\to s+2$.
The double sum is the sum (I) which we have already calculated.
Using
\begin{align}
&\sum_{n=-\infty}^{-1}
\left[(n+\beta)^2  +  \kappa^2\right]^{-s} = \sum_{n=0}^{\infty}
\left[(n+1-\beta)^2  +  \kappa^2\right]^{-s} 
\end{align}
one easily finds for the remaining piece\footnote{
Note that $\zeta_H(0,1)=-1/2$ and $\zeta_H(-2m,1)=0$}
\begin{align}
f_3(s)&=-\sum_{n=0}^{\infty}
\left[(n+1-\beta)^2  +  \kappa^2\right]^{-s} 
\nonumber \\
&= -\kappa^{-2s}
\zeta_H(0,1-\beta)
- \sqrt{\pi}\frac{\Gamma(s-\tfrac{1}{2})}{2\Gamma(s)}
\kappa^{1-2s}\nonumber \\
&-\frac{2\pi^s}{\Gamma(s)} \kappa^{1/2-s}
\sum_{p=1}^{\infty}p^{s-1/2}\cos(2\pi p (1-\beta))
K_{s-1/2}\left(2\pi p \kappa \right) \;.
\end{align}
Differentiating the corresponding contribution to $\zeta(s)$, setting $s=-2$, and 
substituting $\kappa$ yields the Casimir energy,
\begin{align}
\hspace{-7mm}
V_{M}^{\alpha,\beta (II)}
=& V_{M}^{\alpha,\beta (I)} \nonumber \\
&+\frac{4\pi^2}{\mathcal{A}^4L^4\tau_2^2}
\bigg\{ \frac{\mathcal{A}^2 L^4\tau_2^2 M^4}{(4\pi)^4}  
\left[\frac{3}{2}-2\log\left(\frac{M}{\sqrt{\mathcal{A}}\mu_r}\right)\right]
\zeta_H(0,1-\beta)
\nonumber \\
&- \frac{8\pi}{15}  \frac{\mathcal{A}^{5/2}L^5\tau_2^{5/2} M^5}{(4\pi)^5}  \nonumber \\
&+\frac{4}{\pi^2} \left(\frac{(\mathcal{A}\tau_2)^{1/2}L M}{(4\pi)}\right)^{5/2} 
\sum_{p=1}^{\infty}\frac{\cos(2\pi p (1-\beta))}{p^{5/2}}
K_{5/2}\left(2\pi p\frac{\sqrt{\mathcal{A}\tau_2}L M}{4\pi} \right)
\bigg\}.
\end{align}

\subsection{Result}
Putting everything together the Casimir energy can be written as 
\begin{align} 
\hspace{-10mm}
V^{\alpha,\beta}_M 
=&+\frac{M^6 L^2}{3072\pi^3 \mathcal{A}} \left[\frac{11}{12} 
-\log\left(\frac{M}{\sqrt{\mathcal{A}}\mu_r}\right)\right]
\nonumber \\
&- \frac{M^4}{64\pi^2\mathcal{A}^2 }
\left[\frac{3}{4}-\log\left(\frac{M}{\sqrt{\mathcal{A}}\mu_r}\right)\right]
\delta_{\alpha 0}\delta_{\beta 0}
\nonumber \\
&- \frac{M^3 \tau_2^{3/2}}{4 \pi^3\mathcal{A}^{5/2}L} 
\sum_{p=1}^{\infty}\frac{\cos(2\pi p \alpha)}{p^3}
K_{3}\left(p \tfrac{\sqrt{\mathcal{A}}L M}{2\sqrt{\tau_2}}\right) 
 \nonumber \\
&-\frac{32}{\mathcal{A}^4 L^4\tau_2^2} \sum_{p=1}^{\infty} 
 \sum_{m=0}^{\infty}\frac{1}{2^{\delta_{\alpha 0}\delta_{m 0}}}
\frac{\cos(2 \pi p (\beta-(m+\alpha)\tau_1))}{p^{5/2}} 
\left(\tau_2^2(m+\alpha)^2 + \frac{\mathcal{A}L^2 \tau_2 M^2}{(4\pi)^2}  \right)^{\frac{5}{4}} \nonumber \\
& \quad K_{5/2} \left(2\pi \,p \,\sqrt{\tau_2^2(m+\alpha)^2 +  \frac{\mathcal{A}L^2\tau_2 M^2}{(4\pi)^2}   }\right) 
\end{align}
or in terms of the moduli $R_1$, $R_2$ and $\theta$ (and in the frame used in \cite{bcs08})
\begin{align}\label{double}
\hspace{-8mm}
V_M^{\alpha\beta}
=& +\frac{M^6 R_1 R_2}{768\pi}\sin\theta\left(\frac{11}{12} - 
              \log\left(\frac{M}{\mu_r}\right)\right)\nonumber\\
  & - \delta_{\alpha0}\delta_{\beta0}\frac{M^4}{64\pi^2}\left(\frac{3}{4} - 
                \log\left(\frac{M}{\mu_r}\right)\right)\nonumber \\
&- \frac{1}{8\pi^4} \frac{M^3 R_2}{R_1^2}\sin\theta
\sum_{p=1}^{\infty}\frac{\cos(2\pi p \alpha)}{p^3}
K_{3}(\pi p M R_1) \nonumber \\
&-  \frac{2}{\pi^4} \frac{1}{R_2^4} \frac{1}{\sin^4\theta}
\sum_{p=1}^{\infty}
 \sum_{m=0}^{\infty}\frac{1}{2^{\delta_{\alpha0}\delta_{m0}}}
 \frac{\cos(2 \pi p [\beta-(m+\alpha)R_2/R_1 \cos\theta])}{p^{5/2}} 
\nonumber \\ 
& \left(\tfrac{R_2}{R_1}\sin\theta \sqrt{(m+\alpha)^2 +
         \tfrac{M^2R_1^2}{4}}\right)^{5/2}
K_{5/2} \left(2\pi \,p \,\tfrac{R_2}{R_1} \sin\theta
\sqrt{(m+\alpha)^2 + \tfrac{M^2R_1^2}{4}}\right)  \;.
\end{align}
For $\theta=\frac{\pi}{2}$ this agrees with the expression for a rectangular torus \cite{bcs08},
as expected.

\section{Contributions from Vector- and Hypermultiplets}

In this appendix we compare the leading contributions to the Casimir energy from vector and
hypermultiplets, respectively. We will see that the contribution of the vector multiplets 
is generically suppressed compared to the one of the hypermultiplets,
and hence it was justified to neglect this contribution in Eq.~(\ref{casimir}). 

The relative suppression can be seen by an explicit investigation of the mass matrices of 
the gauginos and the hyperscalars, respectively. 
For simplicity we only focus on one single scalar  $\Phi$ and one gaugino $\psi$.
For the present discussion, the relevant part of the 4D Lagrangian reads
\beqra
\mathcal{L} =  -\sum_i \Phi_i M^{s 2}_i \Phi^*_i - \sum_i \psi_i M^{f}_i \chi_i 
+\frac{\lambda \mu^2}{\Lambda^2} \sum_{i j} \Phi_i C^{\Phi}_{i j} \Phi^*_j 
- \frac{h \mu}{2 \Lambda^2} \sum_{i j} 
\psi_i C^{\psi}_{i j} \psi_j
\label{lagrangian}
\eeqra
where for the mode expansion we used the notation
\begin{align}
\Phi(x,y) = \sum_i \Phi_i(x) \xi_i(y)\;, \quad
\int \text{d}^2y \; \xi_i(y)\xi_j(y) = \delta_{ij}\;,
\end{align}
with Kaluza-Klein mass $M_i^{s}$ and 
\begin{equation}
C^{\Phi}_{ij} = \xi_i(0)\xi_j(0)\;.
\end{equation}
In the fermionic case the notation is analogous.
Here $\chi$ is the Weyl fermion which, together with the gaugino, forms the four-component spinor of 
the six-dimensional vector multiplet. 
For simplicity we did not consider any mass terms coming from gauge symmetry breaking, 
although to include also these terms would be straightforward. The first two terms in 
Eq.~(\ref{lagrangian}) follow directly from the KK mode expansion and dimensional reduction, 
whereas the last two terms come from supersymmetry breaking with $\mu$ the supersymmetry breaking mass.
From Eq.~(\ref{lagrangian}) one can read off the scalar as well as the fermionic mass matrix squared.
The fermionic mass matrix squared reads explicitly (in the basis ($\psi,\chi$))
\beq 
\mathcal{M}^{f 2}= 
\left( \begin{array}{cc}
M_i^{f 2} \delta_{i k} + \left(\frac{h}{2 \Lambda^2}\right)^2\sum_{j}C^{\psi}_{i j}C^{\psi}_{j k}\mu^2 & 
\,\,\frac{h}{2 \Lambda^2} C^{\psi}_{j k} \mu M^f_k \\
&\\
\frac{h}{2 \Lambda^2} C^{\psi}_{j k} \mu M^f_k & \,\, M^{f 2}_i \delta_{i k}  
\end{array} 
\right)\;.
\eeq  

Both, the scalars and the gauginos give a contribution to the Casimir energy which 
is proportional to the Trace-Log operator $\tr \log(k^2 + \mathcal{M}^{2})$.
One can decompose the matrix $\mathcal{M}^{2}$ as the sum of two terms  
$\mathcal{M}^{2} =  \mathcal{M}^{2}_0 + \Delta\mathcal{M}^{2}$ (diagonal plus corrections). 
In the fermionic case this reads
\beq 
\mathcal{M}^{f 2}_0= 
\left( \begin{array}{cc}
M_i^{f 2} \delta_{i k} & 0 \\
&\\
 0 & \,\, M^{f 2}_i \delta_{i k}  
\end{array}
 \right) \,\,; 
\Delta \mathcal{M}^{f 2} = 
\left( \begin{array}{cc} 
\left(\frac{h}{2 \Lambda^2} \right)^2 \sum_{j}C^{\psi}_{i j}C^{\psi}_{j k} \mu^2& \,\, 
\frac{h}{2 \Lambda^2}C^{\psi}_{j k} \mu M^f_k \\
&\\
\frac{h}{2 \Lambda^2}C^{\psi}_{j k} \mu M^f_k & 0  
\end{array} 
 \right)
\eeq  
Expanding the Trace-Log operator in powers of $\Delta\mathcal{M}^{2}/(k^2 + \mathcal{M}^{2}_0)$
leads to the expression
\beqra
\tr \log(k^2 + \mathcal{M}^{2}) &=& \tr \log(k^2 + \mathcal{M}^{2}_0) + 
\tr\left(\frac{1}{k^2+\mathcal{M}^{2}_0}\Delta\mathcal{M}^{2}\right) \nonumber\\
&-&\frac{1}{2} \tr\left(\frac{1}{k^2+\mathcal{M}^{2}_0}\Delta\mathcal{M}^{2}
\frac{1}{k^2+\mathcal{M}^{2}_0}\Delta\mathcal{M}^{2} \right) + \dots
\eeqra
which in the fermionic case reads
\beqra
\tr \log(k^2 + \mathcal{M}^{f 2}) &=& \tr \log(k^2 + \mathcal{M}^{f 2}_0) 
+ \left(\frac{h}{2 \Lambda^2}\right)^2
\sum_{i} \frac{1}{k^2  
+ M^{f 2}_i} \sum_k  C^{\psi}_{ik} C^{\psi}_{ki} \mu^2 \nonumber\\
&-& \frac{1}{2}\sum_{i j}  \frac{1}{k^2 + M^{f 2}_i}\frac{1}{k^2 + M^{f 2}_j} 
\left(\frac{h}{2 \Lambda^2} \right)^2
\sum_k C^{\psi}_{i  k}C^{\psi}_{k j} \mu^2 M^{f}_i M^{f}_j + \dots \nonumber \\
\eeqra
Performing the analogous steps in the scalar case we obtain
\beqra
\tr \log(k^2 + \mathcal{M}^{s 2}) &=& \tr \log(k^2 + \mathcal{M}^{s 2}_0) 
+ \frac{\lambda}{\Lambda^2} \sum_{i} \frac{1}{k^2  
+ M^{s 2}_i} C^{\Phi}_{ii} \mu^2 \nonumber\\
&-& \frac{1}{2}\sum_{i j}  \frac{1}{k^2 + M^{s 2}_i}\frac{1}{k^2 
+ M^{s 2}_j}\left(\frac{\lambda}{\Lambda^2}\right)^2
\sum_k C^{\Phi}_{i  k}C^{\Phi}_{k j} \mu^4 + \dots
\eeqra
Since $C^{\psi,\Phi} \sim 1/\mathcal{A}L^2$, we see that the leading 
contribution to the Casimir energy from the vector multiplet
is generically volume and cutoff suppressed compared to the one from the hypermultiplets.


\begin{thebibliography}{10}

\bibitem{wit85}
  E.~Witten,
  Nucl.\ Phys.\  B {\bf 258} (1985) 75.

\bibitem{rat07}
  For recent reviews and references, see\\
  H.~P.~Nilles, S.~Ramos-Sanchez, M.~Ratz and P.~K.~S.~Vaudrevange,
  Eur.\ Phys.\ J.\  C {\bf 59} (2009) 249
  [0806.3905 [hep-th]];
  S.~Raby,
  Eur.\ Phys.\ J.\  C {\bf 59} (2009) 223
  [0807.4921 [hep-ph]].

\bibitem{abc03}
  T.~Asaka, W.~Buchmuller and L.~Covi,
  Phys.\ Lett.\  B {\bf 563} (2003) 209
  [hep-ph/0304142].

\bibitem{bks05}
  W.~Buchmuller, J.~Kersten and K.~Schmidt-Hoberg,
  JHEP {\bf 0602} (2006) 069
  [hep-ph/0512152].

\bibitem{bls07}
  W.~Buchmuller, C.~Ludeling and J.~Schmidt,
  JHEP {\bf 0709} (2007) 113
  [0707.1651 [hep-ph]].

\bibitem{bs08}
  W.~Buchmuller and J.~Schmidt,
  Nucl.\ Phys.\  B {\bf 807} (2009) 265
  [0807.1046 [hep-th]].

\bibitem{bcs08}
W.~Buchmuller, R.~Catena and K.~Schmidt-Hoberg,
  Nucl.\ Phys.\  B {\bf 804} (2008) 70
  [0803.4501 [hep-ph]].

\bibitem{ac83}
  T.~Appelquist and A.~Chodos,
  Phys.\ Rev.\  D {\bf 28} (1983) 772.

\bibitem{pp01}
  E.~Ponton and E.~Poppitz,
  JHEP {\bf 0106} (2001) 019
  [hep-ph/0105021].

\bibitem{pp03}
  M.~Peloso and E.~Poppitz,
  Phys.\ Rev.\  D {\bf 68} (2003) 125009
  [hep-ph/0307379].

\bibitem{ghx05}
  D.~M.~Ghilencea, D.~Hoover, C.~P.~Burgess and F.~Quevedo,
  JHEP {\bf 0509} (2005) 050
  [hep-th/0506164].

\bibitem{gh05}
  G.~von Gersdorff and A.~Hebecker,
  Nucl.\ Phys.\  B {\bf 720} (2005) 211
  [hep-th/0504002];
  C.~Gross and A.~Hebecker,
  0812.4267 [hep-ph].

\bibitem{bht06}
  A.~P.~Braun, A.~Hebecker and M.~Trapletti,
  JHEP {\bf 0702} (2007) 015
  [hep-th/0611102].

\bibitem{hml08}
  H.~M.~Lee,
  JHEP {\bf 0805} (2008) 028
  [0803.2683 [hep-th]].

\bibitem{kks99}
  D.~E.~Kaplan, G.~D.~Kribs and M.~Schmaltz,
  Phys.\ Rev.\  D {\bf 62} (2000) 035010
  [hep-ph/9911293].

\bibitem{clx99}
  Z.~Chacko, M.~A.~Luty, A.~E.~Nelson and E.~Ponton,
  JHEP {\bf 0001} (2000) 003
  [hep-ph/9911323].


\bibitem{bms09}
 W.~Buchm\"uller, J.~M\"oller and J.~Schmidt, DESY 09-026 

\bibitem{knx06}
  T.~Kobayashi, H.~P.~Nilles, F.~Ploger, S.~Raby and M.~Ratz,
  Nucl.\ Phys.\  B {\bf 768} (2007) 135
  [hep-ph/0611020];
  T.~Araki et al., 
  Nucl.\ Phys.\  B {\bf 805} (2008) 124
  [0805.0207 [hep-th]].



\bibitem{knx08}
  R.~Kappl et al.,
  0812.2120 [hep-th].

\bibitem{din99}
For a discussion and references, see for instance\\
  M.~Dine,
  Prog.\ Theor.\ Phys.\ Suppl.\  {\bf 134} (1999) 1
  [hep-th/9903212];
  M.~Dine, G.~Festuccia and A.~Morisse,
  0809.2238 [hep-th].

\bibitem{spa91}
  M.~Spalinski,
  Phys.\ Lett.\  B {\bf 275} (1992) 47;
  J.~Erler, D.~Jungnickel and H.~P.~Nilles,
  Phys.\ Lett.\  B {\bf 276} (1992) 303.

\bibitem{ran77}
R.~A.~Rankin, {\it Modular Forms and Functions}, Cambridge University Press,
1977

\bibitem{Chacko:1999hg}
  Z.~Chacko, M.~Luty and E.~Ponton, 
  JHEP {\bf 07} (2000), 036
  [hep-ph/9909248]

\bibitem{bhk05}
  W.~Buchmuller, K.~Hamaguchi and J.~Kersten,
  Phys.\ Lett.\  B {\bf 632} (2006) 366
  [hep-ph/0506105].

\bibitem{eli94}
  E.~Elizalde,
  J.\ Math.\ Phys.\  {\bf 35} (1994) 6100.


\end{thebibliography}
\end{document}